\documentclass[twocolumn,showpacs,showkeys,preprintnumbers,pra,smaller,
superscriptaddress]{revtex4}

\usepackage{amssymb,amsmath,graphicx,dcolumn,bm,float} 
\usepackage[utf8]{inputenc}
\usepackage[T1]{fontenc}
\usepackage[colorlinks,linkcolor=black,citecolor=blue]{hyperref}

\begin{document}

\title{Frequency locking and controllable chaos through exceptional point in optomechanics}
\author{P. Djorwe}
\email{philippe.djorwe@univ-lille1.fr}
\affiliation{Institut d’Electronique, de Microélectronique et Nanotechnologie, UMR
CNRS 8520 Université de Lille, Sciences et technologies, Villeneuve d’ Ascq 59652, France}

\author{Y. Pennec}
\email{yan.pennec@@univ-lille1.fr}
\affiliation{Institut d’Electronique, de Microélectronique et Nanotechnologie, UMR
CNRS 8520 Université de Lille, Sciences et technologies, Villeneuve d’ Ascq 59652, France}

\author{B. Djafari-Rouhani}
\email{bahram.djafari-rouhani@@univ-lille1.fr}
\affiliation{Institut d’Electronique, de Microélectronique et Nanotechnologie, UMR
CNRS 8520 Université de Lille, Sciences et technologies, Villeneuve d’ Ascq 59652, France}

\begin{abstract}
We engineer mechanical gain (loss) in system formed by two
optomechanical cavities (\textbf{OMCs}), that are mechanically coupled. The gain (loss) is
controlled by driving the resonator with laser
that is blue (red) detuned. We predict analytically the existence of multiple exceptional points (\textbf{EPs}),
  a form of degeneracy where the eigenvalues of the system
coalesce. At each \textbf{EP}, phase transition occurs, and the system switches from
weak to strong coupling regimes and vice versa. In the weak coupling regime, the system locks on an intermediate 
frequency, resulting from coalescence at the \textbf{EP}. In strong coupling regime, however, 
two or several mechanical modes are excited depending on system parameters. The mechanical 
resonators exhibit Rabi-oscillations when two mechanical modes are involved, otherwise the interaction 
triggers chaos in strong coupling regime. This chaos is bounded by \textbf{EPs},
making it easily controllable by tuning these degeneracies. Moreover, this chaotic attractor 
shows up for low driving power, compared to what happens when the coupled \textbf{OMCs} 
are both drived in blue sidebands.  This works opens up promising avenues to use \textbf{EPs} 
as a new tool to study collective phenomena (synchronization, locking effects) in nonlinear systems, and to control chaos.
\end{abstract}

\maketitle

\date{\today}


%
\section{Introduction} \label{Intro}

Optomechanical systems provide a promising platform to explore light-matter
interactions for both technological applications and fundamental physics
\cite{[1]}. Through optomechanics, a mechanical resonator can be studied
from quantum ground state \cite{[2]},\cite{[3]},\cite{[4]} to the
amplified regime characterized with large displacements \cite{[5]},\cite{[6]},\cite{[7]}.

At the parametric instability point, where the backaction-induced
mechanical gain overcomes mechanical loss, mechanical self-oscillations start \cite{[8]},
\cite{[9]},\cite{[10]}, and the system enters into a nonlinear regime. This regime
is a prerequisite to study collective phenomena such as synchronization and
frequency locking \cite{[11]},\cite{[12]},\cite{[13]},\cite{[14]},\cite{[15]}.
Such phenomena have practical applications in
rf communication \cite{[16]}, signal-processing \cite{[17]}, clock synchronization \cite{[18]} and
novel computing and memory concepts \cite{[19]}. In \cite{[13]}, two lasers were used to lock two
optomechanical systems, while the all-optical light-mediated locking of three spatially distant
optomechanical oscillators was achieved using
a single laser source in \cite{[15]}.  The threshold of this locking effect as well 
as the mechanism behind it are not well predicted, but occurs spontaneously as the 
driving strength is increasing \cite{[13]}, \cite{[15]}. 
Therefore, predictability and controllability of locking phenomenon 
become relevant.

In optomechanics, it is well-known that strong driving  strength induces
 period doubling and chaos \cite{[20]},\cite{[21]},\cite{[22]}, \cite{[23]}. Chaotic behaviour is useful for generating
random numbers and implementing secret information processing (see \cite{[24]} and references therein).
However, to apply chaos into a secret communication scheme, good controllability and
low-power threshold are required \cite{[25]},\cite{[26]}. Low-driving
threshold chaos has been achieved in \cite{[25]}, using optical PT-symmetry in an
optomechanical system; while controllable chaos with a low-driving
threshold has been investigated in an electro-optomechanical system in \cite{[26]}.
A system that can handle these issues concerning locking phenomenon and chaos, would be a
good benchmark for technological applications based on nonlinear optomechanics.

Here, we investigate a system that provides both controllability and low-power threshold for chaos as
well as the predictability and control of frequency locking phenomenon. The key point of this is the exceptional
points (\textbf{EPs}), a form of degeneracy in gain and loss systems, where the eigenvalues
coalesce and become conjugate complex numbers \cite{[27]}. The proposal system is formed by two
\textbf{OMCs}, that are mechanically coupled.
The gain and loss are created by symmetrically driving  the cavities with blue and red
detuned lasers, respectively.
Interesting counter-intuitive features and intriguing effects such as stopping light \cite{[28]},
loss-induced suppression and revival of lasing, pump-induced lasing death,
and unidirectional invisibility have been observed in the vicinity of \textbf{EPs} 
(see \cite{[29]} and  references therein).
 Owing to these interesting properties of \textbf{EPs}, here  we show that: (i) chaos and multistability vanish at the \textbf{EP} 
 and (ii) frequency locking effect is induced by \textbf{EP}. This dual effect results from the coalescence of modes. 
 Therefore, switching from multimode to single mode scrucially depends on the \textbf{EP}. These results pave a way to
 control chaos and to predict locking effects in large-scale networks of nonlinear systems by exploiting \textbf{EPs}.
This work is organized as follows. In Sec. \ref{MoEq}, the system and the dynamical equations are described. The predictability 
of frequency  locking effect and control of chaos are presented in Sec. \ref{Frch}. Section \ref{Syn} is devoted to investigate
the transient chaotic behaviour and out of phase synchronization, while Sec. \ref{Concl} concludes the work.

\begin{figure*}[tbh]
\begin{center}
\resizebox{0.38\textwidth}{!}{
\includegraphics{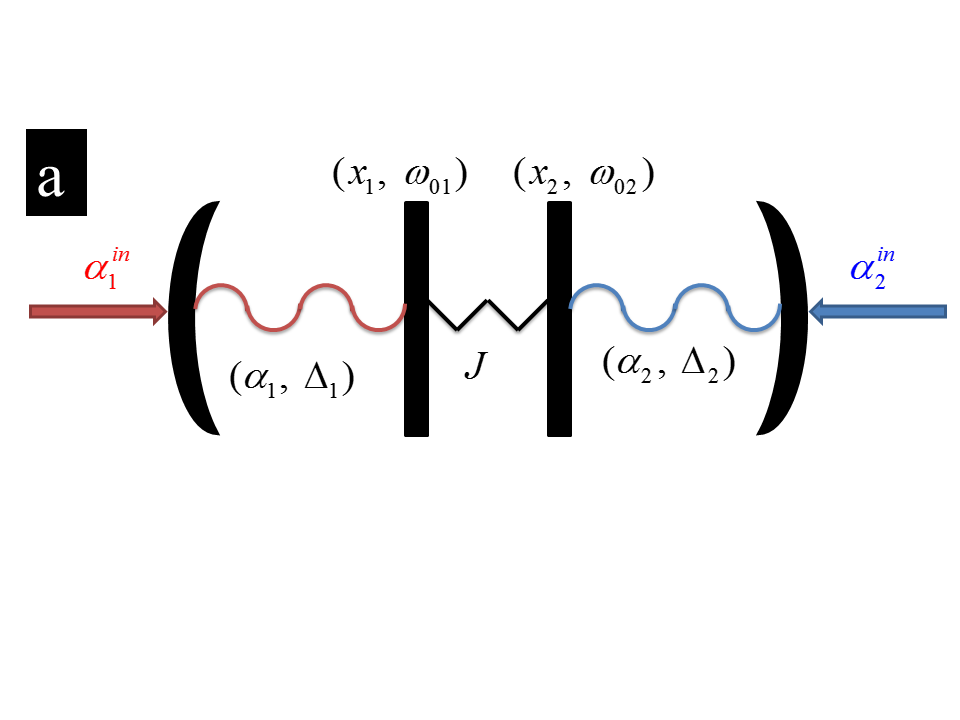}}
\resizebox{0.38\textwidth}{!}{
\includegraphics{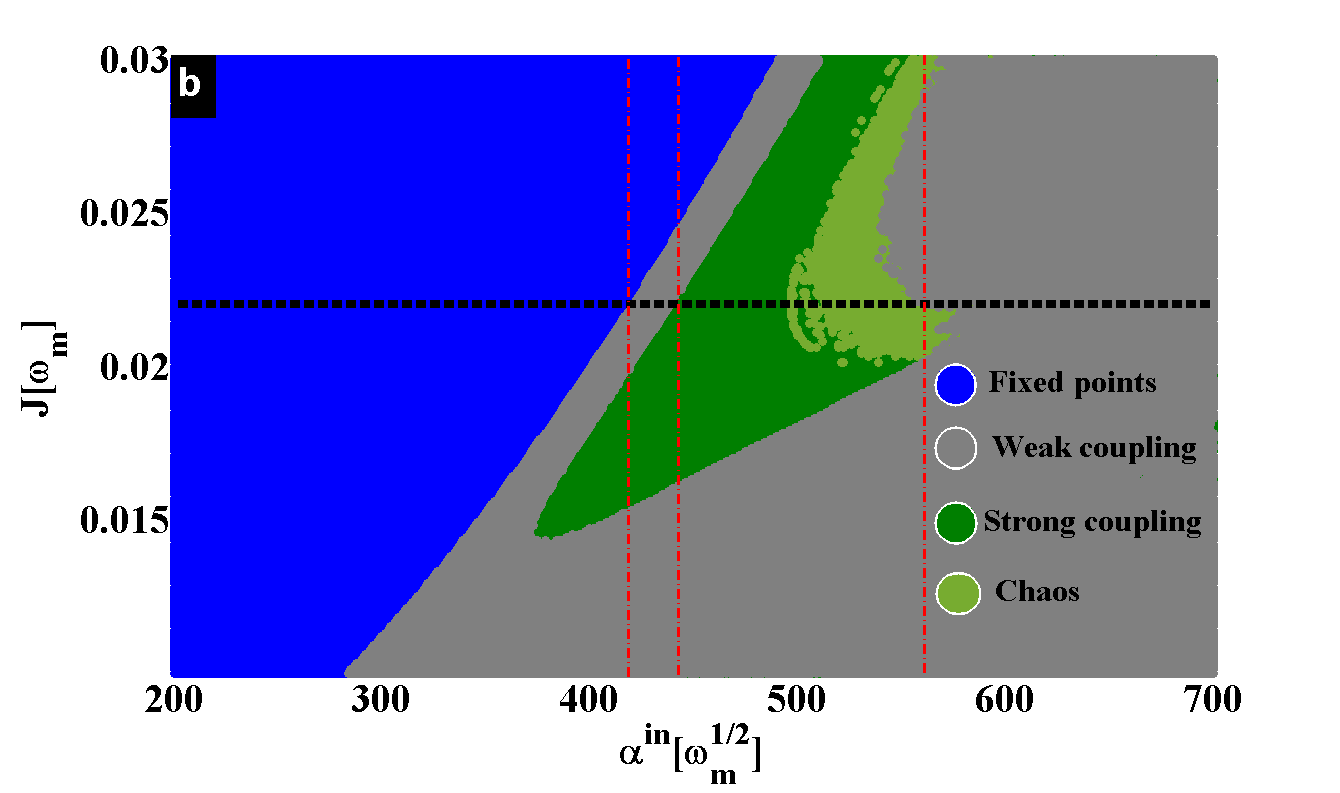}}
\resizebox{0.38\textwidth}{!}{
\includegraphics{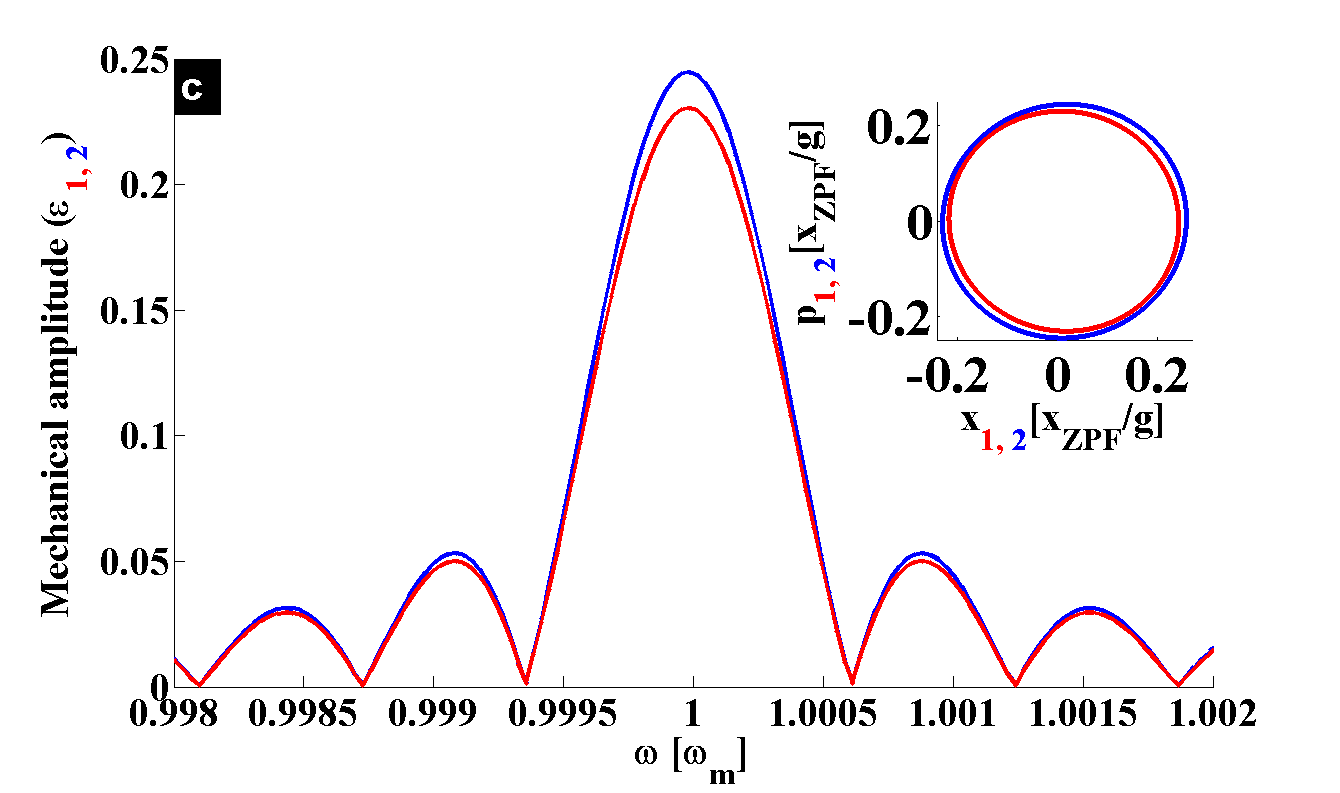}}
\resizebox{0.38\textwidth}{!}{
\includegraphics{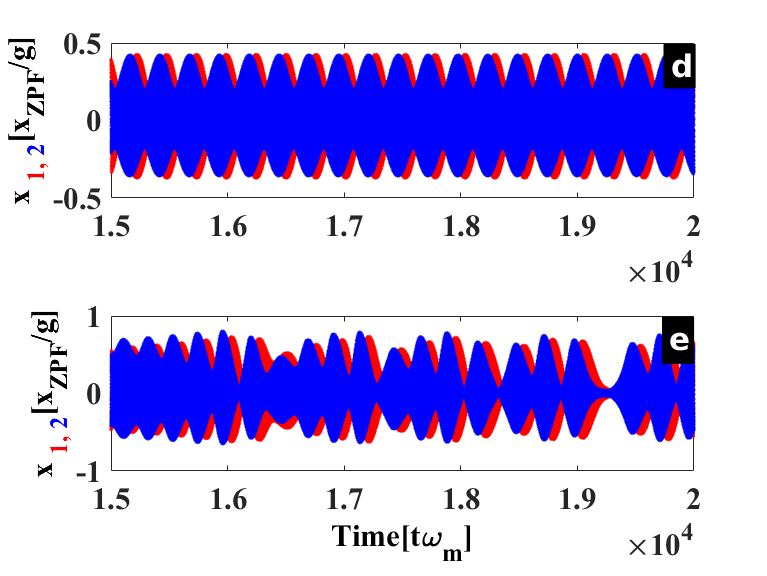}}
\end{center}
\caption{(a) Generic setup. (b) Numerical diagram depicting
the possible regimes involved. (c) Locked frequency and its corresponding phase space representation 
(see inset, $p$ is the momentum). 
(d) and (e) Regular and chaotic Rabi oscillations, respectively. Blue (red) color is related to 
the blue (red) mechanical supermode.
In (c)-(e), $J=2.2\times10^{-2}\omega_m$ and  $\alpha^{in}= (4.3\times10^2, 4.5\times10^2, 5.5\times10^2)\sqrt{\omega_m}$
 respectively.  The other used parameters are, $\gamma_m=10^{-3}\omega_m$, $\kappa=10^{-1}\omega_m$, 
$g=2.5\times10^{-4}\omega_m$, $\omega_{01}=1.002\omega_m$, $\omega_{02}=\omega_m$, $\Delta_1=-\omega_m$ 
and $\Delta_2=\omega_m$.}
\label{fig:Fig1}
\end{figure*}

\section{Modelling and dynamical equations} \label{MoEq}

The system of our proposal is the one in Fig. \ref{fig:Fig1}a, where the cavity labelled $1$ (labelled $2$), is driven with a red (blue)
detuned laser. In the rotating frame of the driving fields, the Hamiltonian ($\hbar=1$) describing 
this system is,
\begin{equation}
H=H_{OM}+H_{int}+H_{drive},
\end{equation}
with 
\begin{equation}
\left\{
\begin{array}
[c]{c}
H_{OM}=\sum_{j=1,2}[-\Delta_{j}a_{j}^{\dag}a_{j}+\omega_{0j}b_{j}
^{\dag}b_{j}-ga_{j}^{\dag}a_{j}(b_{j}^{\dag}+b_{j})]  \\
H_{int}=-J(  b_{1}b_{2}^{\dag}+b_{1}^{\dag}b_{2})  \\
H_{drive}=\sum_{j=1,2}E(a_{j}^{\dag}+a_{j}).
\end{array}
\right. \label{eq01}
\end{equation}
In this Hamiltonian,  $\omega_{0j}$  ($\omega_{01}\neq \omega_{02}$) and  $\Delta_{j}=\omega_{p}^{j}-\omega_{cav}^{j}$ are the
mechanical frequency of the $j^{th}$ resonator and  the optical detuning between the  $j^{th}$ optical
drive ($\omega_{p}^{j}$) and the $j^{th}$ cavity eigenfrequency ($\omega_{cav}^{j}$), respectively. The quantities 
$a_{j}$ and $b_{j}$ are the annihilation  bosonic field operators describing the optical and mechanical resonators, respectively. 
The mechanical displacements $x_{j}$ are connected to $b_{j}$ as $x_{j}=x_{_{\rm{ZPF}}}(b_{j} +b_{j}^{\dag})$, 
where $x_{_{\rm{ZPF}}}$ is the zero-point fluctuation amplitude of the mechanical resonator.
The mechanical coupling strength between the two mechanical resonators is $J$, and  the optomechanical coupling is $g$. 
The amplitude of the driving pump is $E$. The quantum Langevin equations (QLEs) for the operators of the 
optical and the mechanical modes are derived from Eq. (\ref{eq01}) as,
\begin{equation}
\left\{
\begin{array}{c}
\dot{a}_{j}=[i( \Delta_{j}+g(b_{j}^{\dag}+b_{j})) -\frac{\kappa}{2}] a_{j}-i\sqrt{\kappa}(a^{in}+\xi_{a_{j}}), \\
\dot{b}_{j}=-(i\omega_{0j}+\frac{\gamma_{m}}{2}) b_{j}+iJ b_{3-j}+ig a_{j}^{\dag}a_{j}+\sqrt{\gamma_{m}}\xi_{b_{j}},
\end{array}
\right.  \label{eq02}
\end{equation}
where optical ($\kappa$) and mechanical ($\gamma_{m}$) dissipations have been added, and the amplitude of the driving 
pump has been substituted as $E=\sqrt{\kappa}a^{in}$ in order to account for losses. In this form, the input laser power 
$\rm{P_{in}}$ acts through $a^{in}=\sqrt{\frac{\rm{P_{in}}}{\hbar\omega_{p}}}$. The term 
$\xi_{a_{j}}$  ($\xi_{b_{j}}$) denotes the optical  (thermal) Langevin noise at room temperature.

We seek to investigate in the classical limit, where photon and phonon numbers are assumed large in the system, 
and noise terms can be neglected in our analysis. Thus, we rewrite Eq. (\ref{eq02}) into a set of differential equations for the
four  complex scalar fields, \{$\alpha_{j}$\}$_{j=1,2}$ for the optics and \{$\beta_{j}$\}$_{j=1,2}$ for the mechanics, 
standing for the mean values of the operators $\langle a \rangle=\alpha_{j}$ and $\langle b \rangle=\beta_{j}$. 
This leads to the following set of nonlinear equations,
\begin{equation}
\left\{
\begin{array}{c}
\dot{\alpha}_{j}=[i(\Delta_{j}+g(\beta_{j}^{\ast}+\beta
_{j})) -\frac{\kappa }{2}] \alpha_{j}-i\sqrt{\kappa}\alpha^{in}, \\
\dot{\beta}_{j}=-(i\omega_{0j}+\frac{\gamma_{m}}{2}) \beta_{j}+iJ\beta_{3-j}+ig\alpha_{j}^{\ast}\alpha_{j},
\end{array}
\right.  \label{eq1}
\end{equation}
For simplicity, the parameters ($\gamma _{m}$, $g$, $\kappa$) are 
assumed to be degenerated for both cavities. Throughout the work, we assume 
the hierarchy of parameters $\gamma_{m},g\ll \kappa \ll \omega_{0j}$, 
similar to the experiments carried out in the resolved sideband regime \cite{[30]},\cite{[31]}.

In Figs. \ref{fig:Fig1}(b)$-$(e), we show the overall properties of the steady state solutions of Eq. (\ref{eq1}),  
where all the transient dynamics has died out. Three regimes can be identified in ($\alpha^{in}, J$) parameter's space. 
As the driving $\alpha^{in}$ increases for a fixed $J=2.2\times10^{-2}\omega_m$ (see dashed line in Fig. \ref{fig:Fig1}b), 
the system switches from the linearized regime (blue area) to the nonlinear one (gray and green colors) through 
the onset of the self-induced oscillations. In the nonlinear regime, the system switches twice into 
weak coupling regimes (gray color), and once into a strong coupling regime (green color). The meaning of weak (strong) 
coupling regime will be given later on.  Each transition in weak coupling regime is followed by limit cycle oscillations, 
and both mechanical resonators lock and start oscillating with a common frequency 
(see Fig. \ref{fig:Fig1}c). However, this frequency locking 
phenomenon is destroyed when the system jumps into the strong coupling regime, where Rabi oscillations 
show up (see Fig. \ref{fig:Fig1}d and Fig. \ref{fig:Fig1}e). These Rabi oscillations can be 
regular (Fig. \ref{fig:Fig1}d) or chaotic like-behaviour (Fig. \ref{fig:Fig1}e and light green area in Fig. \ref{fig:Fig1}b). 
This chaotic attractor is bounded between two limit cycle regimes. Such phase transitions, 
between weak and strong couplings in coupled gain/loss system, are reminiscent  of \textbf{EP} \cite{[32]},\cite{[33]}. 
It results that, (i) our system features multiple \textbf{EPs} \cite{[34]}, which are useful: (ii) to induce 
frequency locking, (iii) and to control chaotic dynamics.

To get insight of the \textbf{EP} features, we approach the limit cycle oscillations
 by the ansatz, $\beta_{j}(t)=\bar{\beta}_{j}+A_{j}\exp(-i\omega_{_{\rm{lock}}}t)$ \cite{[35]}. 
$\bar{\beta}_{j}$ is a constant shift in the origin of the movement,
 $A_{j}$ is the slowly time dependent amplitude of the cycles, and $\omega_{_{\rm{lock}}}$ is the 
 mechanical locked frequency. Similar to multistability in optomechanics \cite{[35]},
 this ansatz aims to provide analytical tools, describing the feature of multiple \textbf{EPs}.
Using this ansatz, it is straightforward to integrate $\alpha_{j} (t)$ out of the 
full system (see Appendix \ref{App.B}), resulting in 
effective equations of motion for just the mechanical resonators having the form
$i\partial t\Psi =H_{eff}\Psi$. We have set the state vector
$\Psi =\left(\beta_{1},\beta_{2}\right) ^{T}$  and the effective Hamiltonian is,

\begin{equation}
H_{eff}=
\begin{bmatrix}
\omega_{eff}^{1}-i\frac{\gamma_{eff}^{1}}{2} & -J \\
-J & \omega_{eff}^{2}-i\frac{\gamma_{eff}^{2}}{2}
\end{bmatrix}.
\label{eqa}
\end{equation}
This Hamiltonian  has the eigenvalues,

\begin{equation}
\lambda_{\pm}\simeq  \frac{\omega_{eff}^{1}+\omega_{eff}^{2}}{2}-\frac{i}{4}\left(\gamma_{eff}^{1}+\gamma
_{eff}^{2}\right) \pm \frac{\sigma}{2}.  \label{eq2}
\end{equation}
Here $\omega_{eff}^{j}=\omega_{0j}+\delta \omega_{j}$
and $\gamma_{eff}^{j}=\gamma_{m}+\gamma_{opt}^{j}$ are the effective frequencies
and  dampings respectively. The quantity $\sigma \approx\sqrt{4J^{2}-\frac{\Delta
\gamma _{eff}^{2}}{4}}$, with $\Delta\gamma_{eff}=\gamma_{eff}^{2}-\gamma_{eff}^{1}$, is
amplitude dependent through the normalised amplitude 
$\epsilon_{j}=\frac{2g\rm{Re}(A_{j})}{\omega_{_{\rm{lock}}}}$. 
Indeed, the optical dampings $\gamma_{opt}^{j}$ are expressed as ,

\begin{equation}
\gamma_{opt}^{j}=\frac{2(g\kappa \alpha ^{in})^{2}}{\epsilon_{j}}
\sum_{n}\frac{J_{n+1}\left(-\epsilon_{j}\right) J_{n}\left(
-\epsilon_{j}\right)}{\left\vert h_{n+1}^{j\ast}h_{n}^{j}\right\vert
^{2}}\text{,}  \label{eqb}
\end{equation} 
where $J_{n}$ is the Bessel function,  $h_{n}^{j}=i\left(
n\omega_{_{\rm{lock}}}-\tilde{\Delta}_{j}\right) +\frac{\kappa}{2}$ and $\tilde{
\Delta}_{j}=\Delta _{j}+\delta_{j}$, with $\delta_{j} = 2g(\bar{\beta}_{j})$, is the nonlinear detuning 
(see Appendix \ref{App.B}). 
The eigenfrequencies and the dampings of the system are
defined as the real ($\omega_{\pm}=\Re(\lambda_{\pm})$) and imaginary ($\gamma_{\pm}=\Im(\lambda_{\pm})$)
parts  of $\lambda_{\pm}$, respectively. However, the quantities $\omega_{\pm}$ and
$\gamma_{\pm}$ depend on $\sigma$, delimiting the weak and strong regimes aforementioned in Fig. \ref{fig:Fig1}b. 
The strong coupling regime is defined for $J>\frac{\Delta \gamma _{eff}}{4}$, while the weak coupling 
one holds for $J<\frac{\Delta \gamma _{eff}}{4}$. 
The \textbf{EPs}, phase transitions between these two regimes, are defined by 
$J=\frac{\Delta \gamma _{eff}}{4}$. This induces $\sigma=0$, whose solutions predict  multiple 
 \textbf{EPs}  \cite{[34]}, owing to the oscillating nature of $\sigma$
(see the dashed line in Fig. \ref{fig:Fig1}b). After demonstrating the emergence of multiple 
\textbf{EPs}, we take a step further, showing that \textbf{EPs} can be used as a new paradigm both 
for achieving frequency locking and to control chaos.  

\section{Frequency locking and chaos} \label{Frch}

\subsection{Frequency locking}

\begin{figure}[tbh]
\begin{center}
\resizebox{0.45\textwidth}{!}{
\includegraphics{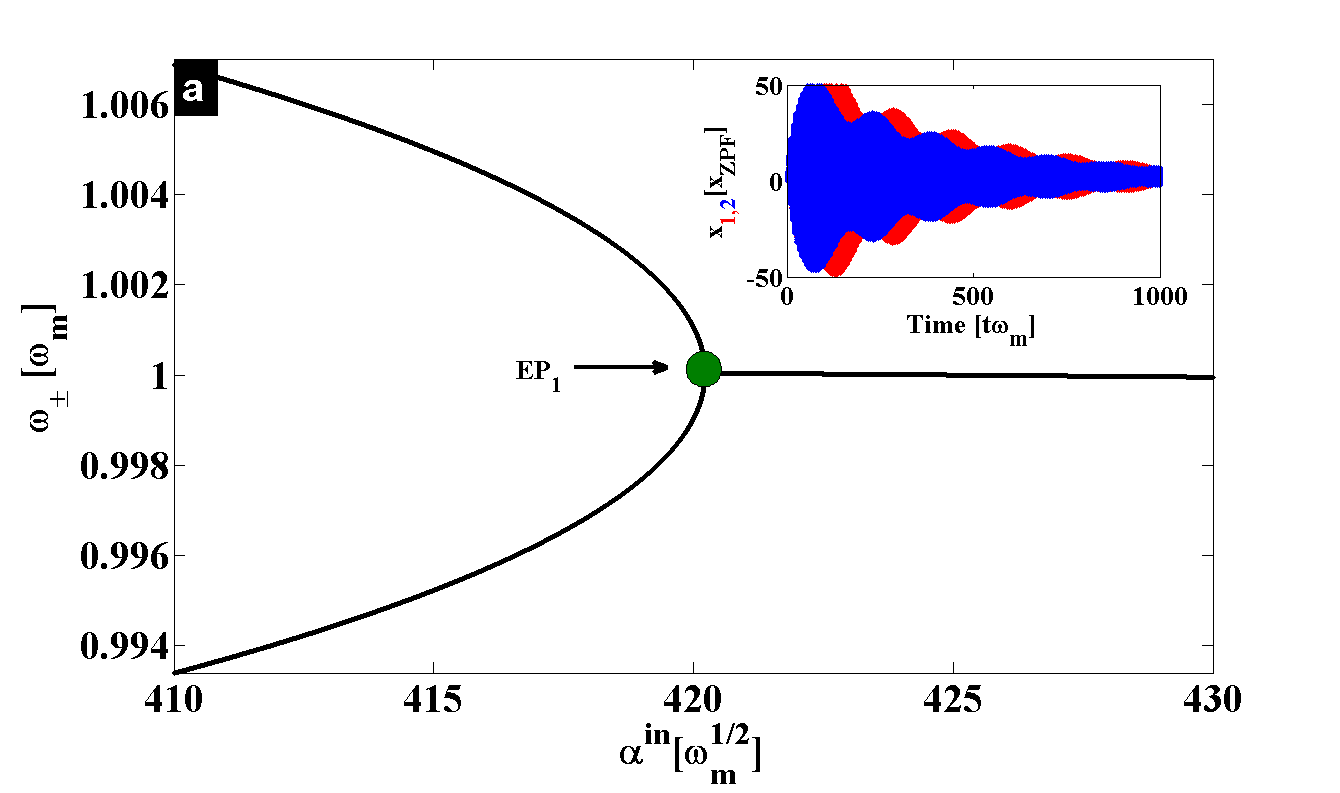}}
\resizebox{0.45\textwidth}{!}{
\includegraphics{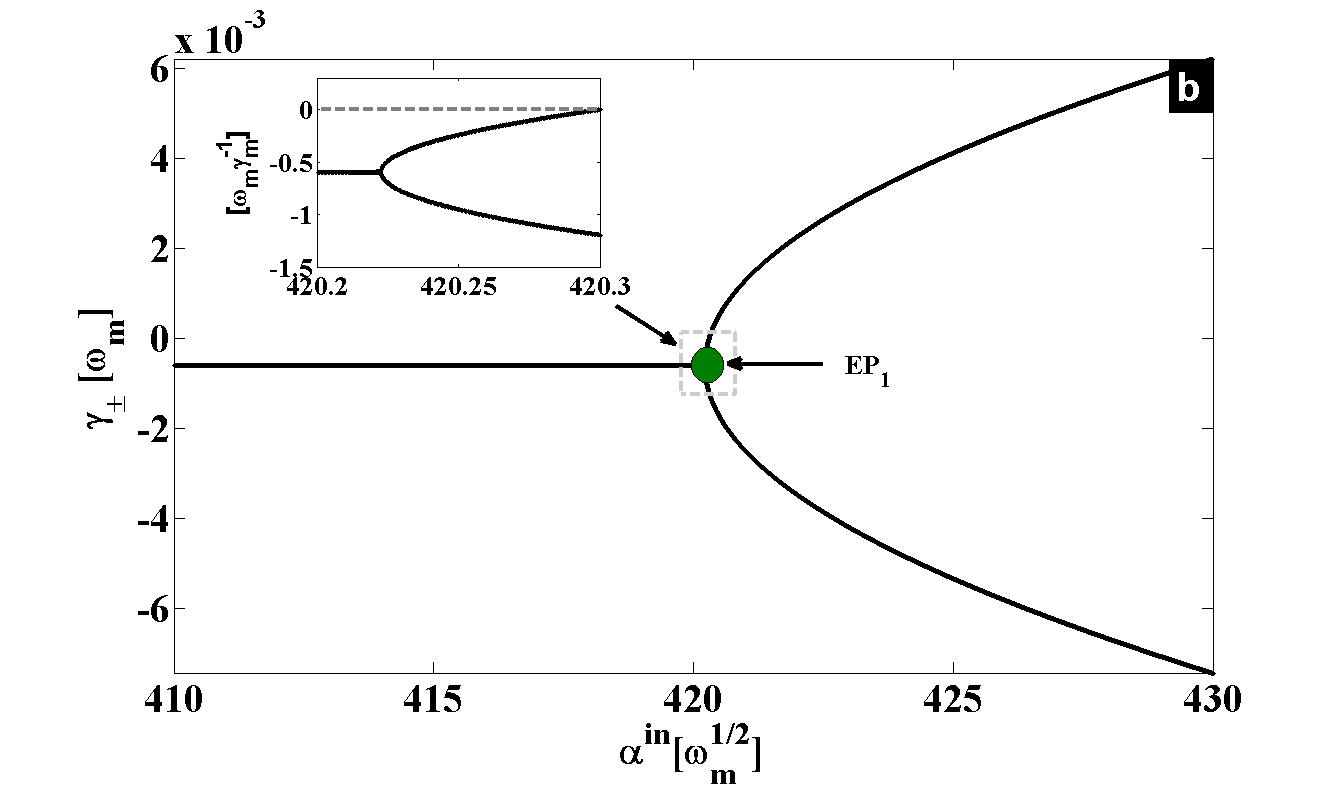}}
\end{center}
\caption{(a), (b) Real and imaginary part of the eigenmodes, respectively. 
Inset of (a) shows (mechanical) Rabi oscillations at $\alpha^{in}=3\times10^2\sqrt{\omega_m}$. 
The coupling strength is $J=2.2\times10^{-2}\omega_m$ and the other parameters are as in Fig. \ref{fig:Fig1}.}
\label{fig:Fig2}
\end{figure}

In the linear regime, $\epsilon_{j}\rightarrow 0$, we use \{$\sigma(\epsilon_{j})$,
$\mu(\epsilon_{j})$\} $\rightarrow$ \{$\sigma_{0}$, $\mu(0)$\}, where $\mu$ stands for 
any amplitude-dependent term. The mechanical resonators oscillate with
two eigenfrequencies (see Fig. \ref{fig:Fig2}a), 
\begin{equation}
\omega_{\pm}\simeq  \frac{\omega _{eff}^{1}(0)+\omega _{eff}^{2}(0)}{2} \pm \frac{\sigma_{0}}{2},  \label{eq3}
\end{equation}
that exchange energy through Rabi oscillations \cite{[32]},\cite{[33]} as depicted in the inset of Fig. \ref{fig:Fig2}a.
 These Rabi oscillations have  an exponentially decaying profile whose 
 is defined by the imaginary parts of the eigenvalues (see Fig. \ref{fig:Fig2}b),

\begin{equation}
\gamma_{\pm}=\frac{-\left(\gamma _{eff}^{1}(0)+\gamma
_{eff}^{2}(0)\right)}{4}.  \label{eq4}
\end{equation}
The quantity $\Delta\gamma_{eff}$ is quadratic in $\alpha ^{in}$
(see Eq. (\ref{eqb})), and it will overcome $4J^{2}$ as $\alpha ^{in}$ is increasing, that is at
 \textbf{EP}$_{1}$.  Consequently, the two mechanical
resonators spontaneously lock at the frequency
$\omega_{_{\rm{lock}}}= \frac{\omega _{eff}^{1}+\omega _{eff}^{2}}{2}$, as it can be deduced from Eq. (\ref{eq3}).
This locking effect persists until another \textbf{EP} is reached. This constitutes one of our findings, showing that
frequency locking is achieved through \textbf{EP}. This result opens up novel prospects for 
applications of \textbf{EPs} in realizing locking modes
in optomechanics and others similar devices. 

\subsection{Chaos}

\begin{figure*}[tbh]
\begin{center}
\resizebox{0.38\textwidth}{!}{
\includegraphics{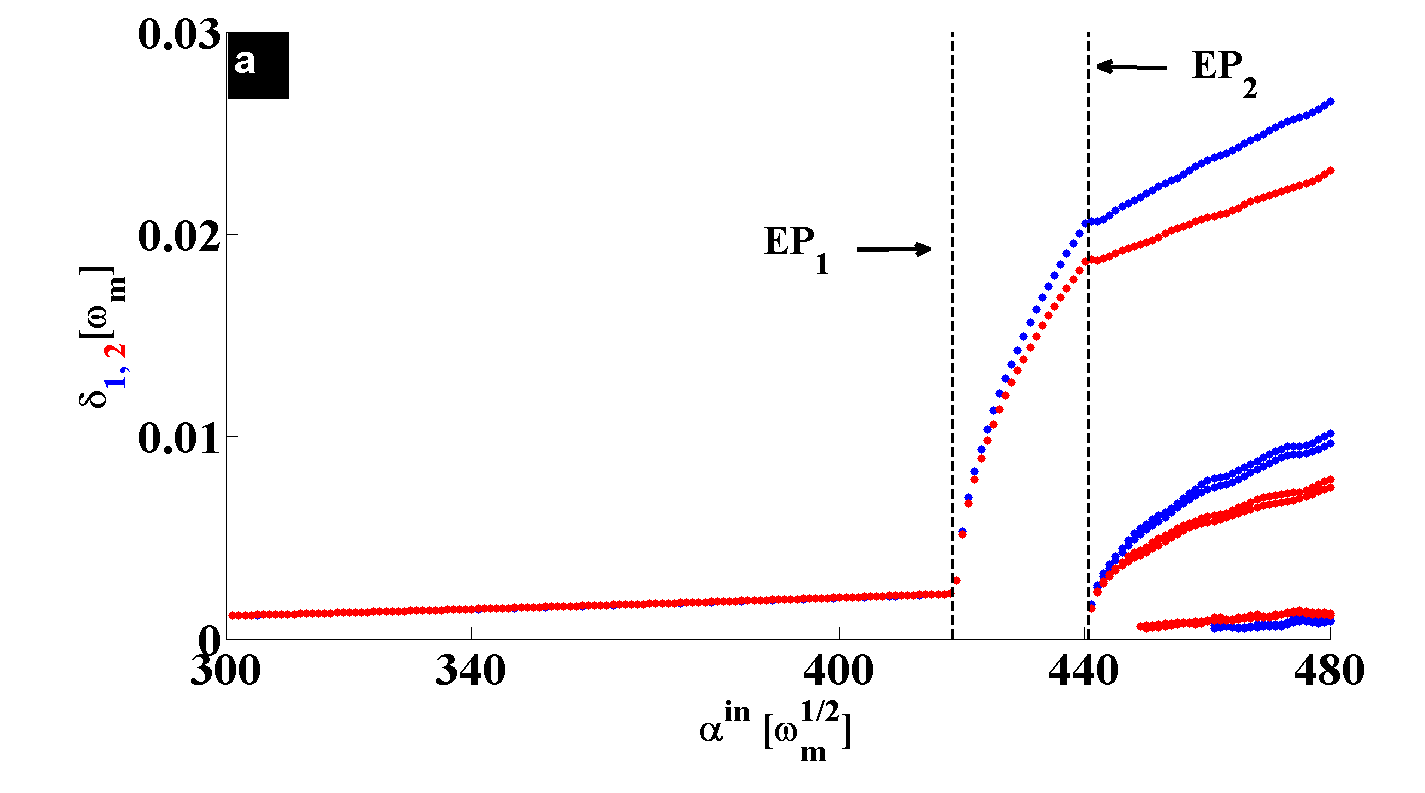}}
\resizebox{0.47\textwidth}{!}{
\includegraphics{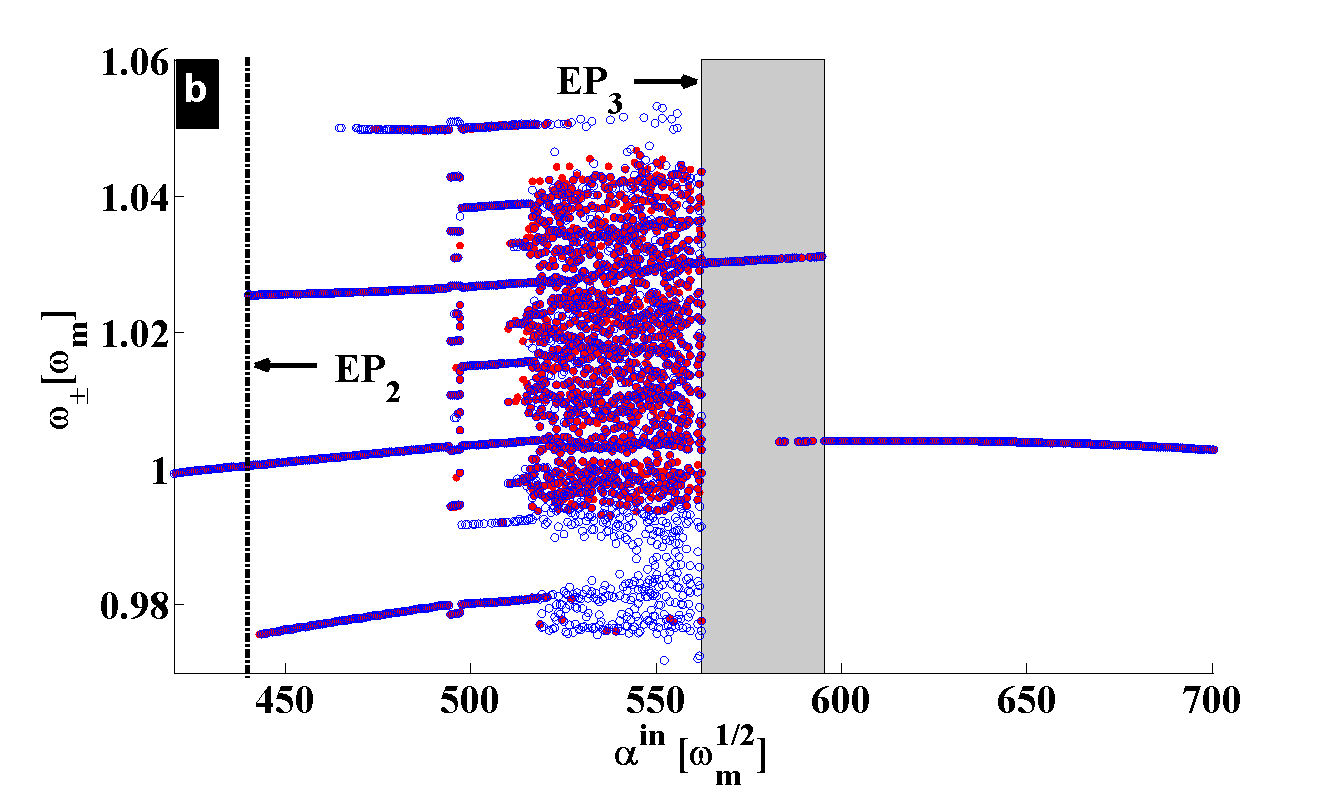}}
\resizebox{0.47\textwidth}{!}{
\includegraphics{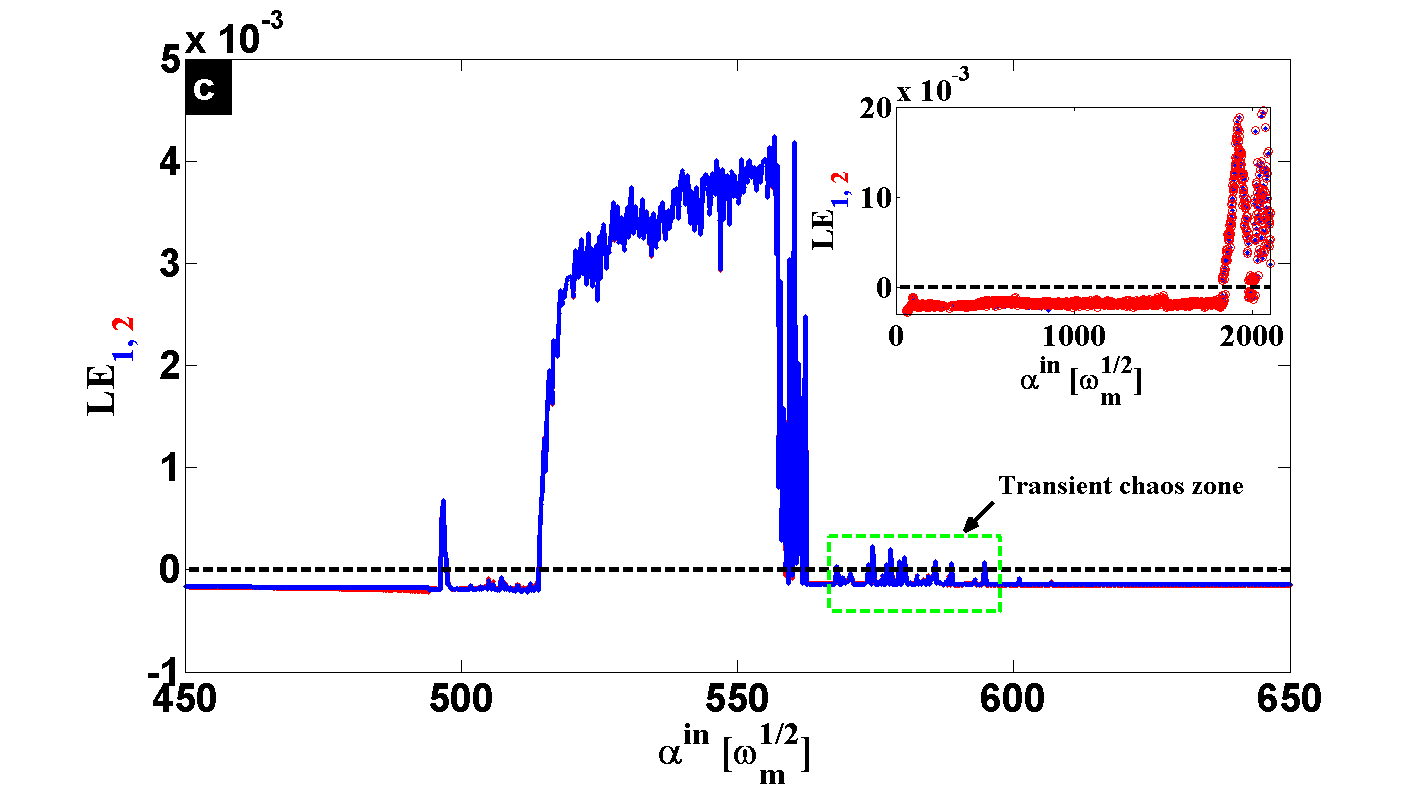}}
\resizebox{0.47\textwidth}{!}{
\includegraphics{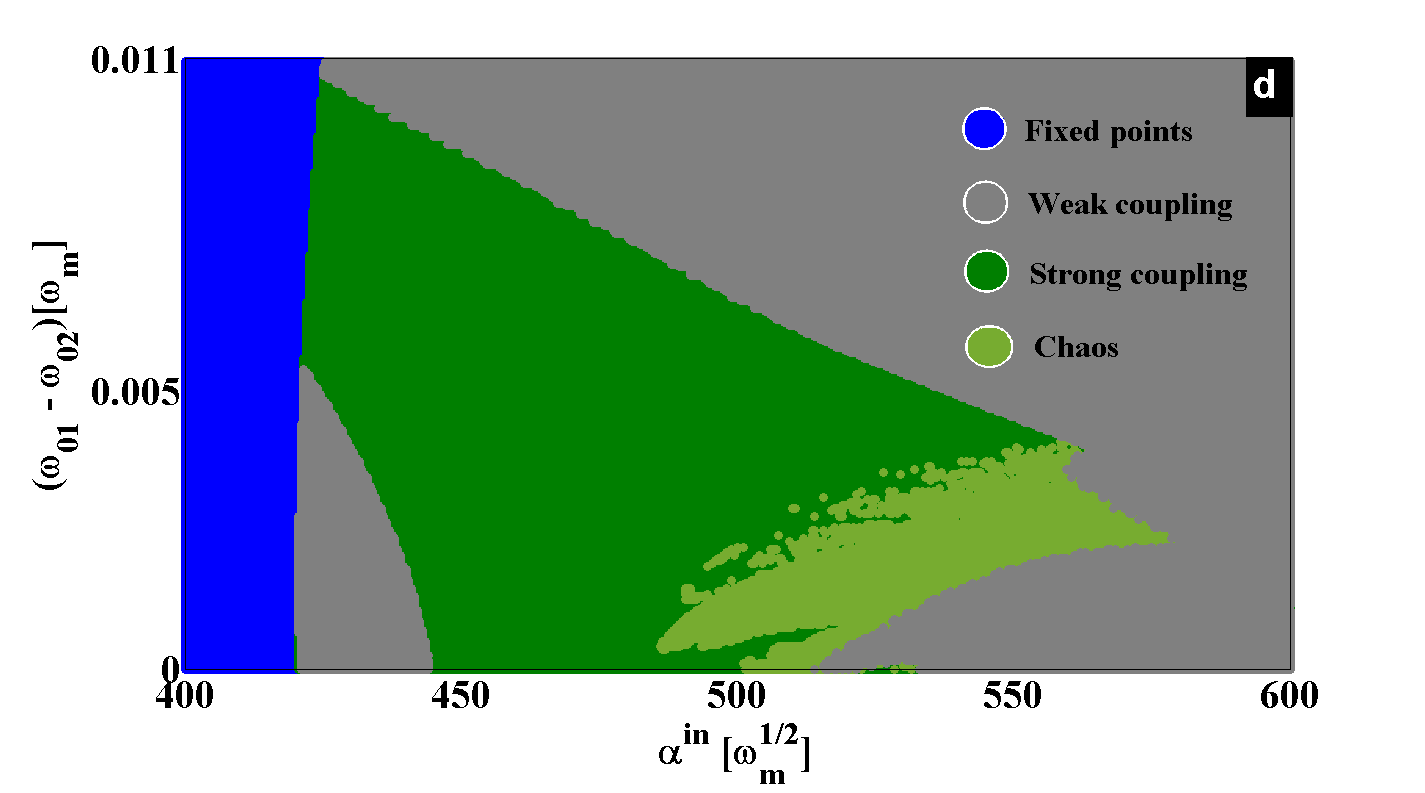}}
\end{center}
\caption{(a) Frequency shift and origin of nonlinearities. 
(b) Frequencies of the mechanical resonators versus $\alpha^{in}$. 
(c) Corresponding Lyapunov Exponent (\textbf{LE}) versus $\alpha^{in}$. The inset of (c) is the \textbf{LE} 
 for the analog blue$-$blue configuration.  (d) Overview of dynamical states related to 
 frequency mismatch $\omega_{01}-\omega_{02}$ at $J=2.2\times10^{-2}\omega_m$. }
\label{fig:Fig3}
\end{figure*}

In the nonlinear regime, linear approximation is preserved for weak amplitudes 
($\epsilon_{j}\ll 1$) \cite{[30]},\cite{[36]}, and both dissipations 
($\gamma_{\pm}$) keep the same sign (see inset of  Fig. \ref{fig:Fig2}b). 
For non-negligible $\epsilon_{j}$, the oscillations of $\sigma$ can
lead to  multiple \textbf{EPs}. For $J=2.2\times10^{-2}\omega_m$ for instance, \textbf{EP}$_{2}$ and 
\textbf{EP}$_{3}$ are induced as shown in Fig. \ref{fig:Fig3}a and  Fig. \ref{fig:Fig3}b (see also vertical and 
horizontal line intersections in Fig. \ref{fig:Fig1}b). 
Weak coupling holds between \textbf{EP}$_{1}$ and \textbf{EP}$_{2}$. 
Beyond \textbf{EP}$_{2}$, the system jumps into a strong 
coupling regime, where Rabi oscillation emerge (see Fig. \ref{fig:Fig1}d).
As the driving increases, the optical nonlinearity $\delta_{j}$ splits into multistable 
solutions (see Fig. \ref{fig:Fig3}a), affecting the frequency of Rabi oscillations $\sigma$. In our proposal, this 
multistability process constitutes a route to chaos \cite{[23]}. When the optomechanical nonlinearities become 
comparable to the optical linewidth ($\delta_{j}\sim \frac{\kappa}{2}$), chaotic oscillations are 
triggered in the system. This can be seen in Fig. \ref{fig:Fig3}b, showing range of frequencies continuum. 
This figure is obtained by collecting peaks and corresponding frequencies, 
of the mechanical steady states, from Fast Fourier Transform (FFT). Such bifurcation diagram in frequency space is useful here, 
since it has the advantage of well-tracking dynamics of Rabi oscillations (see Appendix \ref{App.A}).  
Through Lyapunov Exponent (\textbf{LE}) \cite{[24]},\cite{[25]}, we have confirmed this 
chaotic behaviour in Fig. \ref{fig:Fig3}c. The negative (positive)  value of \textbf{LE} 
indicates that the system exhibits periodic (chaotic) dynamics. For quasi-periodic behaviour, discrete frequencies 
in Fig. \ref{fig:Fig3}b, \textbf{LE} is close to zero. As the driving strength is growing, \textbf{EP}$_{3}$ is reached, and the system 
switches back into the weak coupling regime. Features stemming from the presence of \textbf{EP}$_{3}$ are the disappearance 
of Rabi oscillations \cite{[32]} and the spontaneous emergence of frequency locking 
(see Fig. \ref{fig:Fig3}b for $\alpha^{in}\sim 5.6\times10^{2} \sqrt{\omega_m}$). 
It results that, the chaotic attractor is bounded between \textbf{EP}$_{2}$ and \textbf{EP}$_{3}$.
Threshold of this chaos can be controlled by tuning \textbf{EPs} through system's parameters. We focus our 
investigation here on the mechanical frequency mismatch $\omega_{01}-\omega_{02}$. The reason lies 
on the difficuties to engineer two identical mechanical resonators, that have exactly the same frequencies.
For a fixed $J=2.2\times10^{-2}\omega_m$,  Fig. \ref{fig:Fig3}d shows that, the mechanical resonators can be no longer 
strongly coupled if their frequency mismatch exceeds $(\omega_{01}-\omega_{02})\gtrsim10^{-2}\omega_m$. 
Furthermore, large frequency mismatch destroys chaotic dynamics, since chaos is limited for  
$(\omega_{01}-\omega_{02})\lesssim5\times10^{-3}\omega_m$ as shown by the light green color in Fig. \ref{fig:Fig3}d.
It follows that, an increase (decrease) of the mechanical frequency mismatch controls (induces) chaotic dynamics.
Conversely, increasing $(\omega_{01}-\omega_{02})$ enhances frequency locking effect.  
This provides a method of manipulating and controlling chaos through \textbf{EPs}, making it
useful in large technological platforms \cite{[23]}. This is our second finding, suggesting a 
bounded and controllable chaos through \textbf{EPs} in coupled \textbf{OMCs}. The main ingredient for the emergence of this chaos is
a strong coupling between the mechanical resonators, instead
of being a strong driving strength \cite{[20]},\cite{[21]},\cite{[22]}, \cite{[23]}. 

In the above discussion, we considered the blue-red configuration of coupled cavities. 
For a matter of comparison, the inset of Fig. \ref{fig:Fig3}c shows the \textbf{LE} obtained in the analog blue-blue configuration, using 
the same parameters. It results that, the threshold of chaos is reduced almost four times in our proposal (see also \cite{[24]}, \cite{[25]}).
Lowering threshold of chaos is a requiring element in a secret communication scheme,  and our work provides a new paradigm based on tunability
of \textbf{EP}.

\section{Transient chaos and $\pi$-synchronization} \label{Syn}
\subsection{Transient chaos}

\begin{figure*}[tbh]
\begin{center}
\resizebox{0.42\textwidth}{!}{
\includegraphics{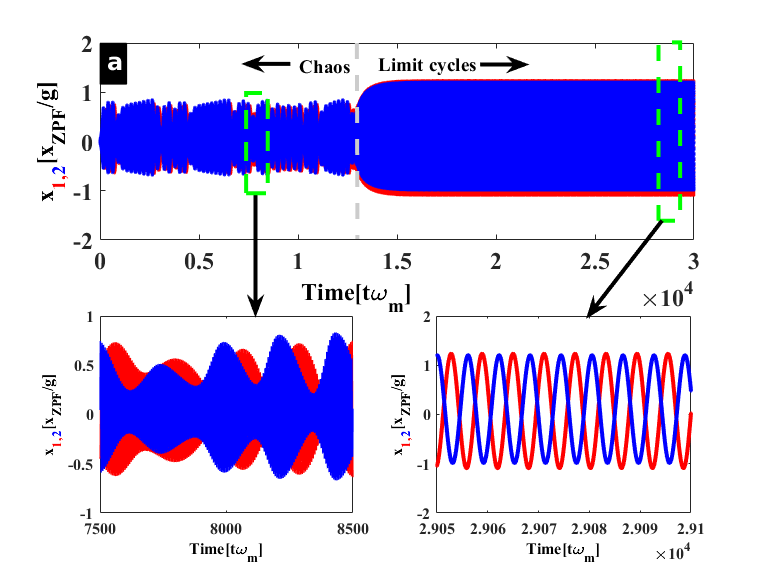}}
\resizebox{0.45\textwidth}{!}{
\includegraphics{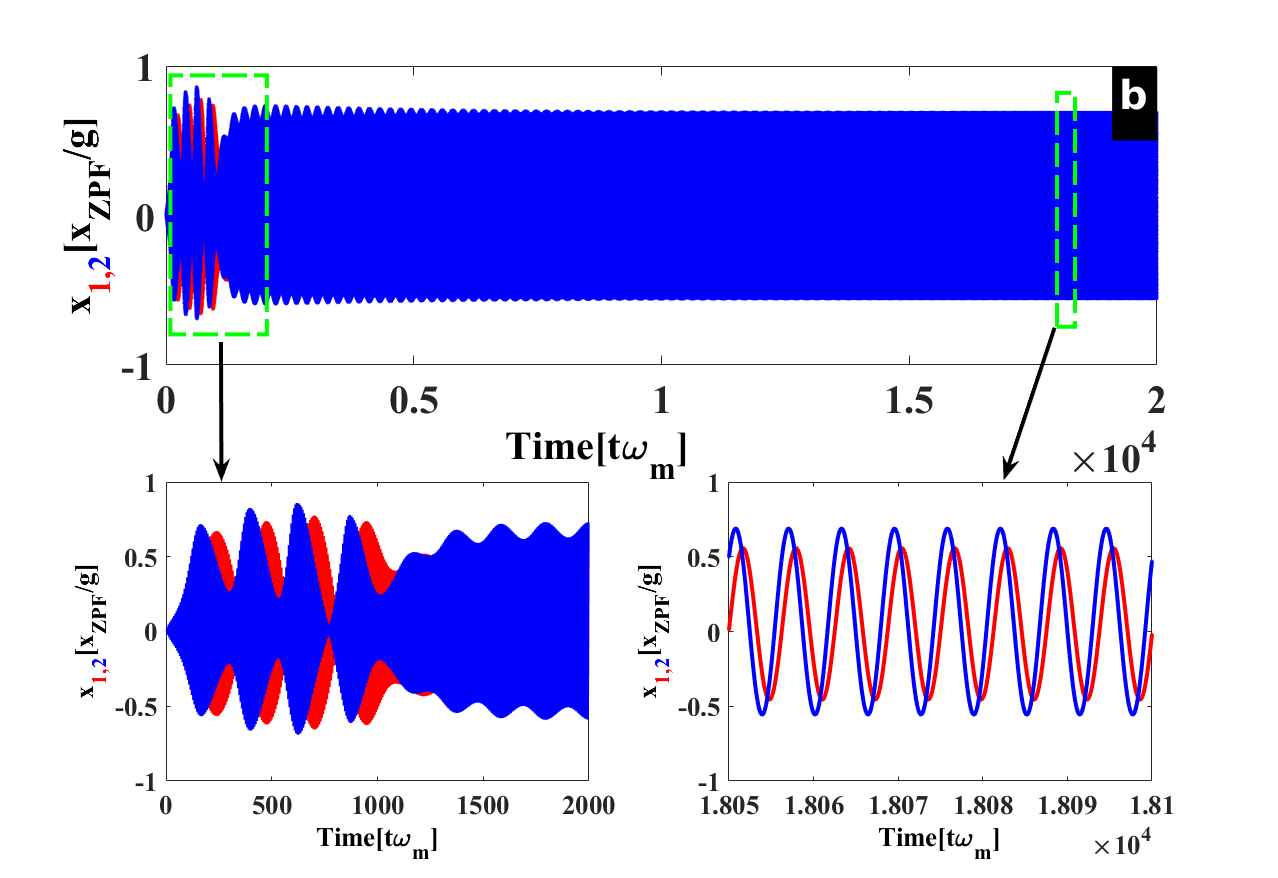}}
\end{center}
\caption{(a) Transient chaos induced $\pi$-synchronization at $\alpha^{in}=5.7\times10^{2}\sqrt{\omega_m}$. 
(b) Frequency locking with transient Rabi-oscillations at $J=5.9\times10^{-3}\omega_m$. (a)-(b) correspond 
to the coupling strength $J=2.2\times10^{-2}\omega_m$.}
\label{fig:Fig4}
\end{figure*}

The phenomenon of transient chaos was recently studied in optomechanics \cite{[39]}. Besides
being a physically meaningful phenomenon by itself,  these authors have shown that transient chaos constitutes
a bridge for the quantum-classical transition. However, we show here that transient chaos induces 
transition towards frequency locking. At the \textbf{EP}$_{3}$, chaotic dynamics vanishes, 
and the system locks back at $\omega_{_{\rm{lock}}}$ (see Fig. \ref{fig:Fig3}b). This locked state depends on 
whether there is a coexistence between transient chaos and limit cycle attractors or not. In the former 
case, this coexistence locks the system on a higher energy state, otherwise the locking is achieved 
on the lower energy state. As we can see in the gray area in Fig. \ref{fig:Fig3}b, the system starts on the 
upper branch (higher energy), and gradually switches on the lower branch (lower energy) as the driving strength 
is increasing. This is depicted in Fig. \ref{fig:Fig4}, where we have chosen two values of $\alpha^{in}$, 
one for the upper branch (see Fig. \ref{fig:Fig4}a) and the other on the lower branch (see Fig. \ref{fig:Fig4}b). 
It results that, transient chaos and limit cycle attractors coexist on the upper branch, while 
Rabi-oscillations precede limit cycles on the lower branch. When transient chaos is involved, the \textbf{LE} 
starts diverging, and decays over time in order to match the appropriate limit cycle dynamics. However, 
this relaxation time is long that the chaotic signature persists in the \textbf{LE} 
(see the green box in Fig. \ref{fig:Fig3}c). As the configuration in Fig. \ref{fig:Fig4}b is 
more stable than the one in Fig. \ref{fig:Fig4}a \cite{[39]}, after a short competition between upper and lower branches, 
the system finally settles into the lower branch that is more stable. Besides its dynamical 
aspect, the transient state determines the kind of collective phenomenon exhibited by the 
final steady state (see zooms  in Fig. \ref{fig:Fig4}a and Fig. \ref{fig:Fig4}b).

\subsection{$\pi$-synchronization}
\begin{figure}[tbh]
\begin{center}
\resizebox{0.5\textwidth}{!}{
\includegraphics{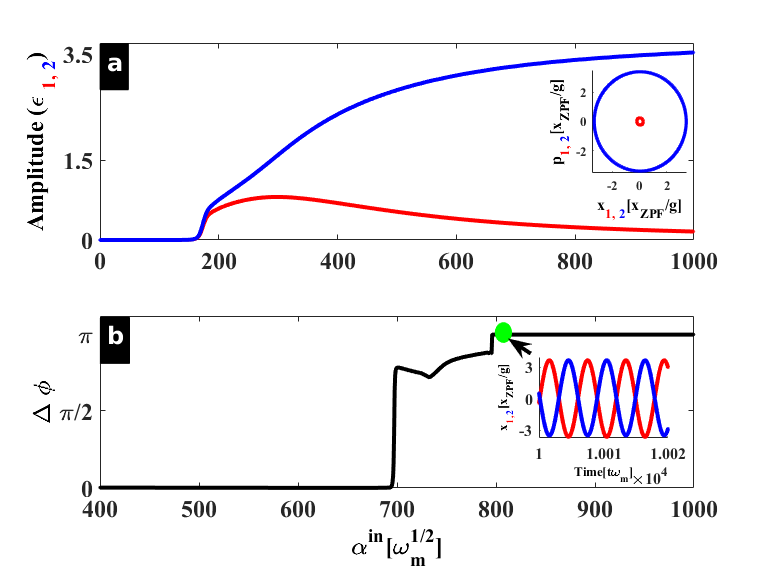}}
\end{center}
\caption{(a) Asymmetric dissipation at $J=5\times10^{-3}\omega_m$.
(b) Standard deviation of $\delta\phi$ showing $\pi$-synchronization for $J=6\times10^{-2}\omega_m$. 
The insets in (a) and (b) show limit cycle and time propagation at $\alpha^{in}=8\times10^{2}\sqrt{\omega_m}$. 
The other parameters are as in Fig. \ref{fig:Fig1}.}
\label{fig:Fig5}
\end{figure}

Beyond this transient regime (gray area in Fig. \ref{fig:Fig3}b), only the locked state with the lower energy persists and 
the mechanical resonators exhibit two different behaviours based on their phase difference. Either they oscillate out of 
phase or they exhibit limit cycle oscillations with different dissipations, 
\begin{equation}
\gamma_{\pm }=\frac{-\left(\gamma_{eff}^{1}+\gamma
_{eff}^{2}\right)}{4}\pm \frac{\sigma}{2}.  \label{eq7}
\end{equation}
In the latter case, the asymmetry between the dissipation rates ($|\gamma_{-}| \neq |\gamma_{+}|$) induces 
unidirectional flow of phonons \cite{[37]}, \cite{[38]} between the resonators  as shown in Fig. \ref{fig:Fig5}a. 
However, the first case happens when $\gamma_{eff}^{1}\sim -\gamma_{eff}^{2}$, leading to 
$\gamma_{\pm }\sim\pm \frac{\sigma}{2}$ \cite{[33]}. Consequently, the mechanical resonators 
carry out approximately broken $\mathcal{PT-}$symmetry dynamics, resulting in a $\pi-$synchronization 
($\Delta\phi=\phi_{1}-\phi_{2}=\pi$) as a signature. This is shown in Fig. \ref{fig:Fig5}b,
where the standard deviation of $\Delta\phi$ is represented. The phase is defined as being 
$\tan(\phi_{j})=\frac{\rm{Im}(\beta_{j})}{\rm{Re}(\beta_{j})}$, confirming the phase difference of
$\pi$ when $\gamma_{\pm }\sim\pm \frac{\sigma}{2}$.  
  
The parameters we have used here are similar to those in the recent experiments \cite{[30]}, \cite{[31]}. This
 offers the prospects to experimentally reproduce the present results. Moreover, this study can be extended to
 a wide variety of  optomechanical systems, including hybrid optical-microwave setups and electromechanical
 systems. Our findings do not necessarily need  $\mathcal{PT-}$symmetry, and the resonators 
 can have different frequencies (see Fig. \ref{fig:Fig3}d).

\section{Conclusion} \label{Concl}
In conclusion, we have studied two optomechanical systems that are mechanically coupled. By
driving the cavities, one by blue detuned laser and the other with a red
detuned laser, we have respectively created gain and loss on these mechanical resonators. We  have predicted 
analytically, the existence of multiple \textbf{EPs}. The system switches from weak to
strong coupling regimes through these \textbf{EPs}. In the weaks coupling regimes, we demonstrated
frequency locking effect induced by these degeneracies. In the strong coupling regime instead, 
 we have shown that optical nonlinearities trigger chaos. This chaotic attractor is bounded between two \textbf{EPs},
providing an accurate way to control it by adjusting gain/loss parameters. This work offers the prospects
to use \textbf{EPs} as a new tools for controlling and thresholdless chaos. Furthermore, \textbf{EPs} open up a
promising route for realizing collective phenomena (locking effect, synchronization) in nonlinear devices.

\section*{Acknowledgments}

This work was supported by the European Commission FET OPEN H2020
project PHENOMEN-Grant Agreement No. 713450.

\appendix \label{App}

\section{Dynamical states} \label{App.A}

\begin{figure*}[tbh]
\centering
\par
\begin{center}
\resizebox{0.38\textwidth}{!}{
\includegraphics{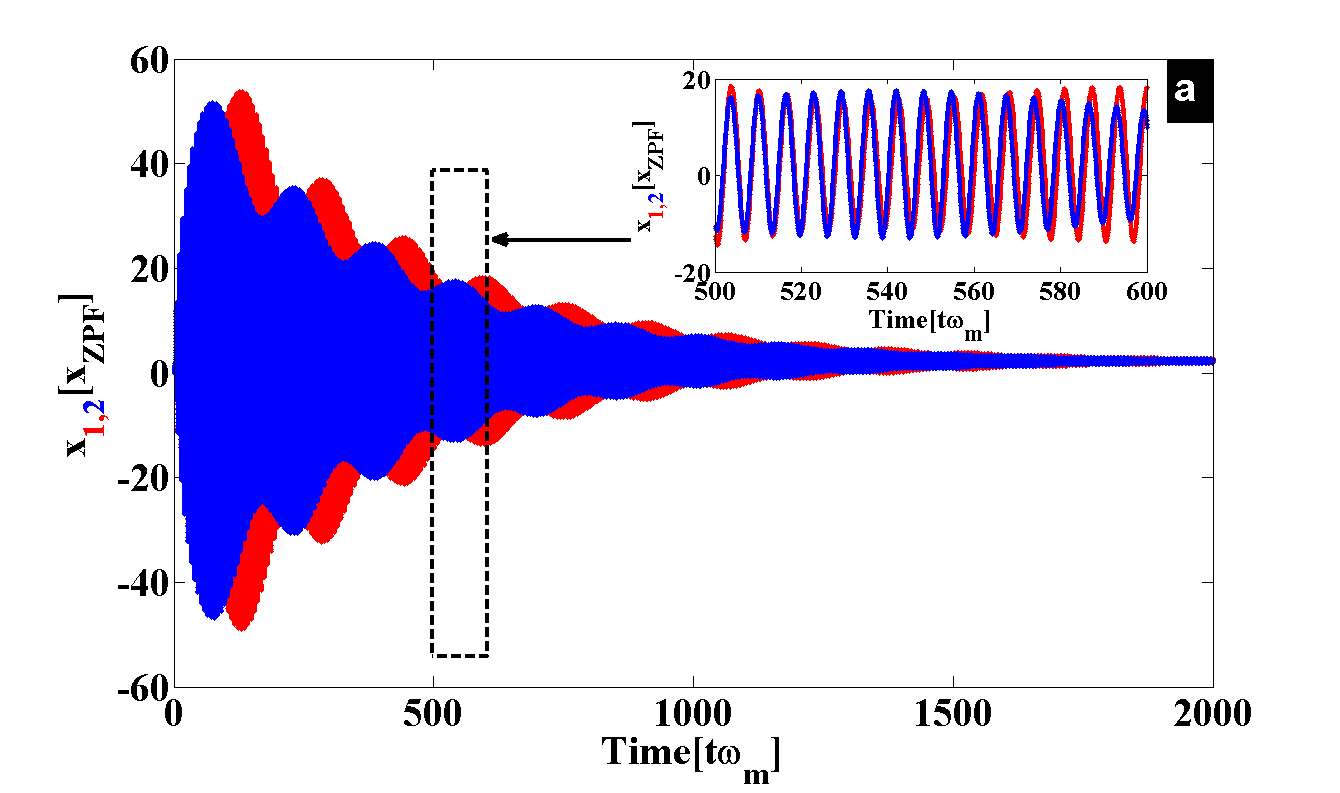}}
\resizebox{0.38\textwidth}{!}{
\includegraphics{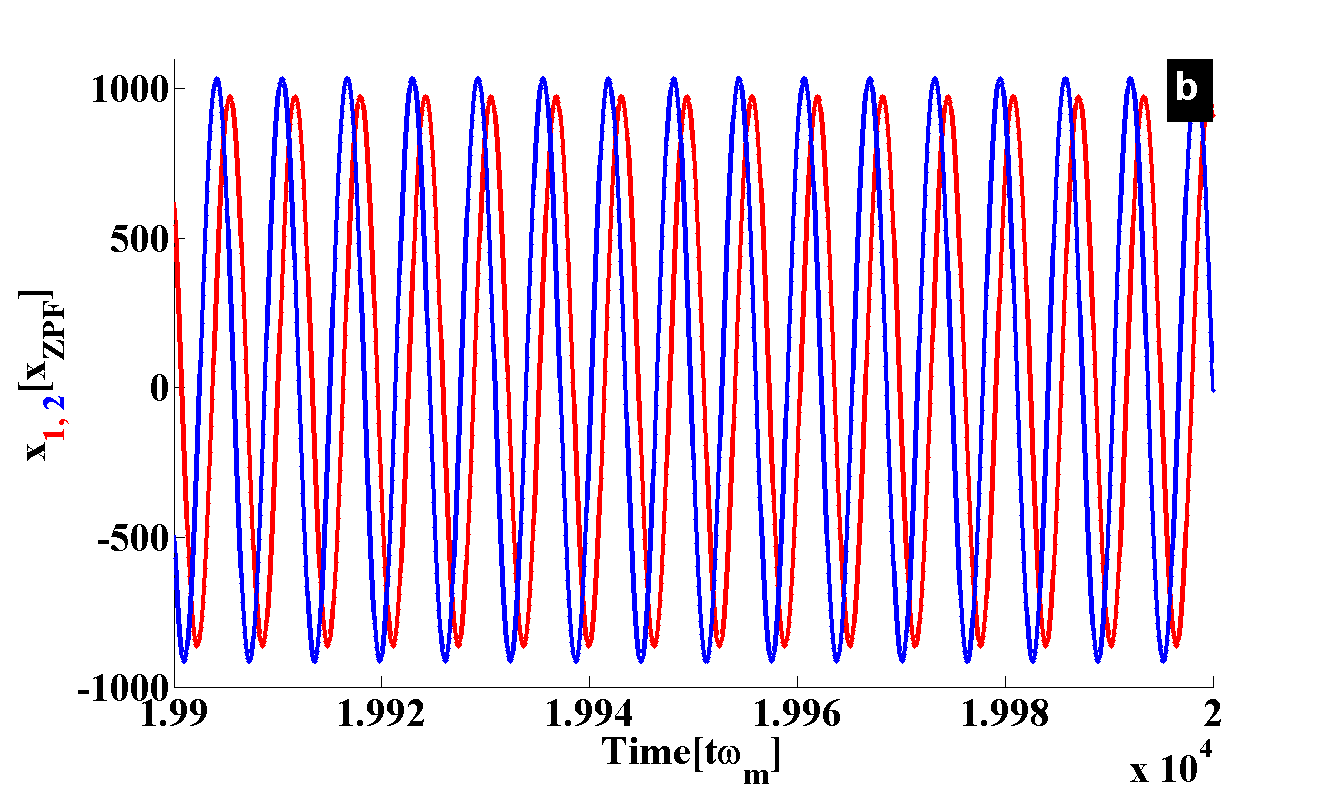}}
\resizebox{0.38\textwidth}{!}{
\includegraphics{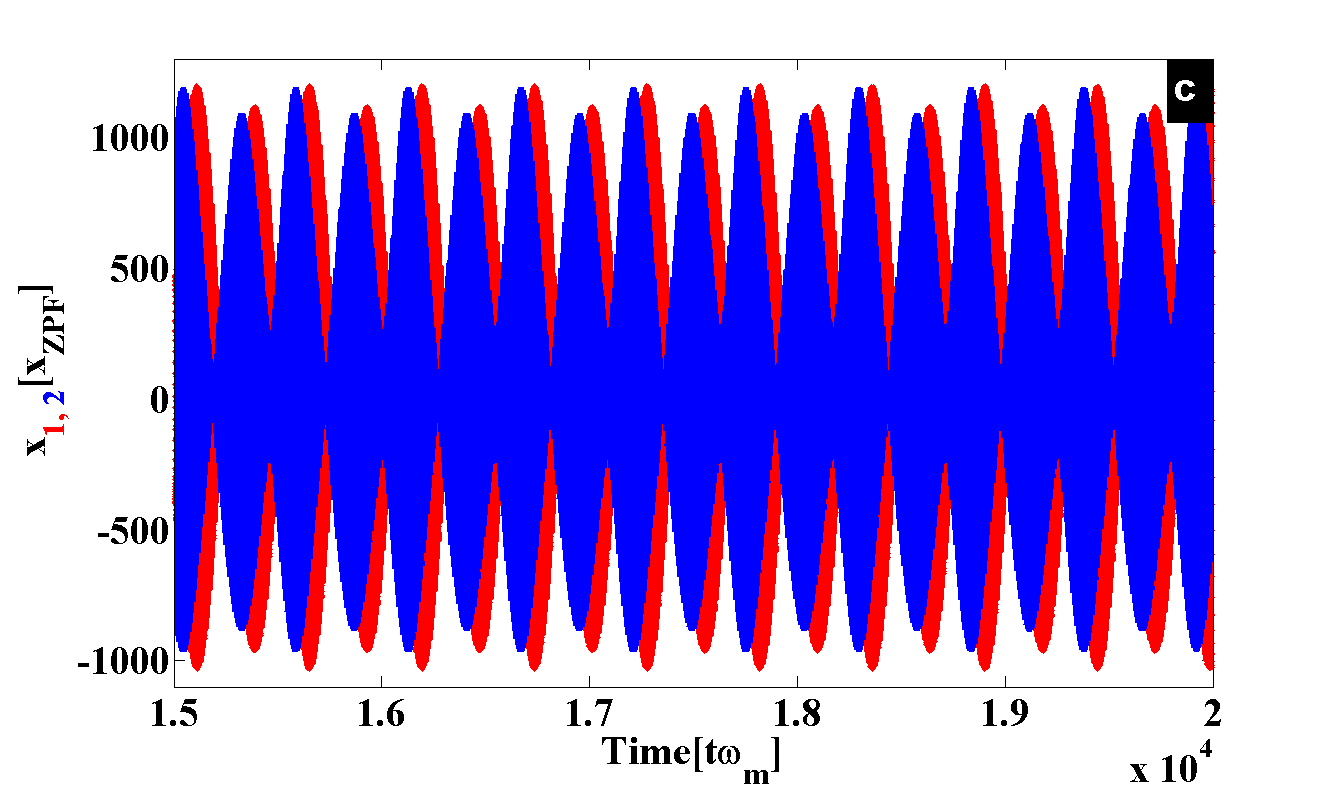}}
\resizebox{0.38\textwidth}{!}{
\includegraphics{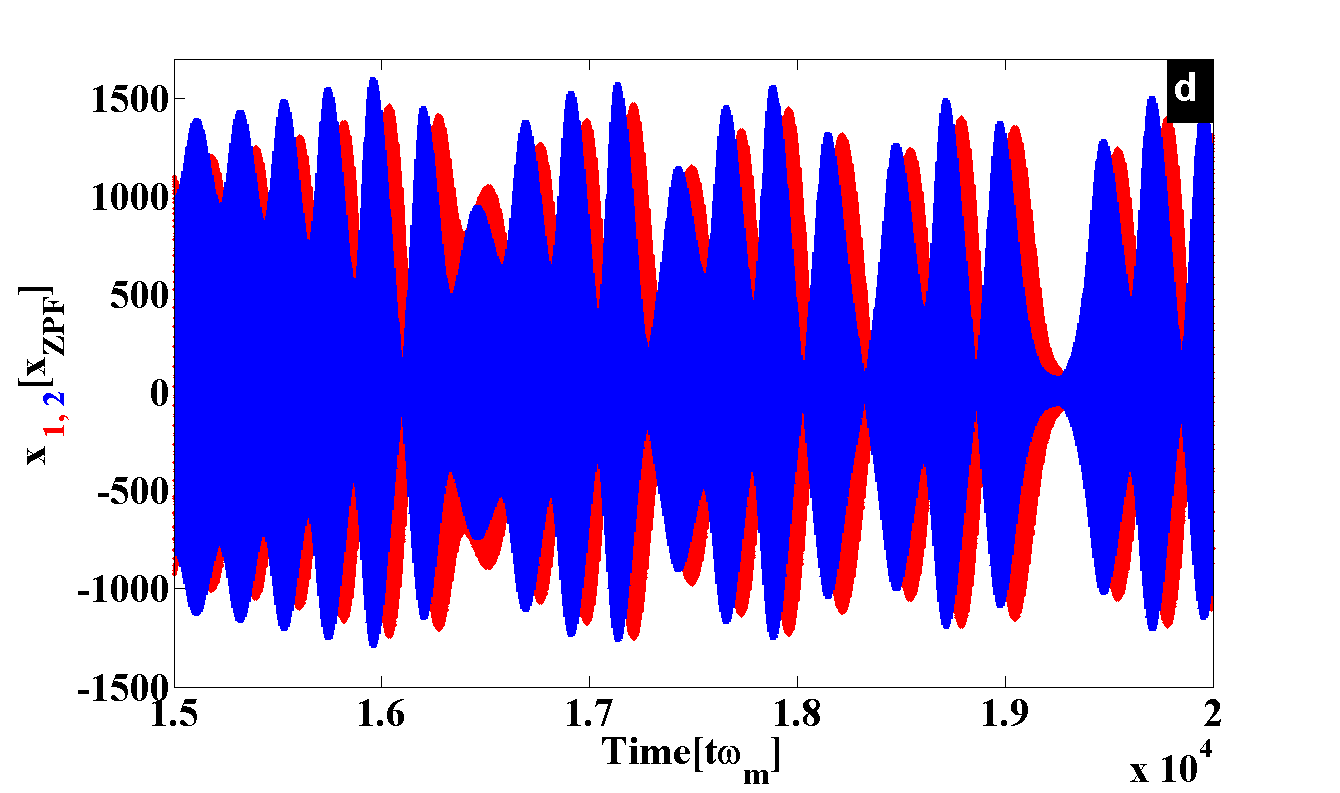}}
\end{center}
\caption{Time propagations.  (a) Fixed point state at $\alpha^{in}= 3\times10^2\sqrt{\omega_m}$. 
(b) Limit cycle state at $\alpha^{in}= 4.3\times10^2\sqrt{\omega_m}$. The corresponding spectrum and phase space trajectory are
those of Fig.1c in the main text. (c) Quasi-periodic state at $\alpha^{in}= 5\times10^2\sqrt{\omega_m}$. 
 (d) Chaotic state at $\alpha^{in}= 5.5\times10^2\sqrt{\omega_m}$. States (c) and (d) can be confirmed from the Lyapunov 
Exponent in the main text. The coupling strength is $J=2.2\times10^{-2}\omega_m$ and the other parameters remain 
the same as in Fig.1 in the main text. Blue (red) color is related to the blue (red) mechanical supermode.}
\label{fig:FigS1}
\end{figure*}

\begin{figure*}[tbh]
\centering
\par
\begin{center}
\resizebox{0.38\textwidth}{!}{
\includegraphics{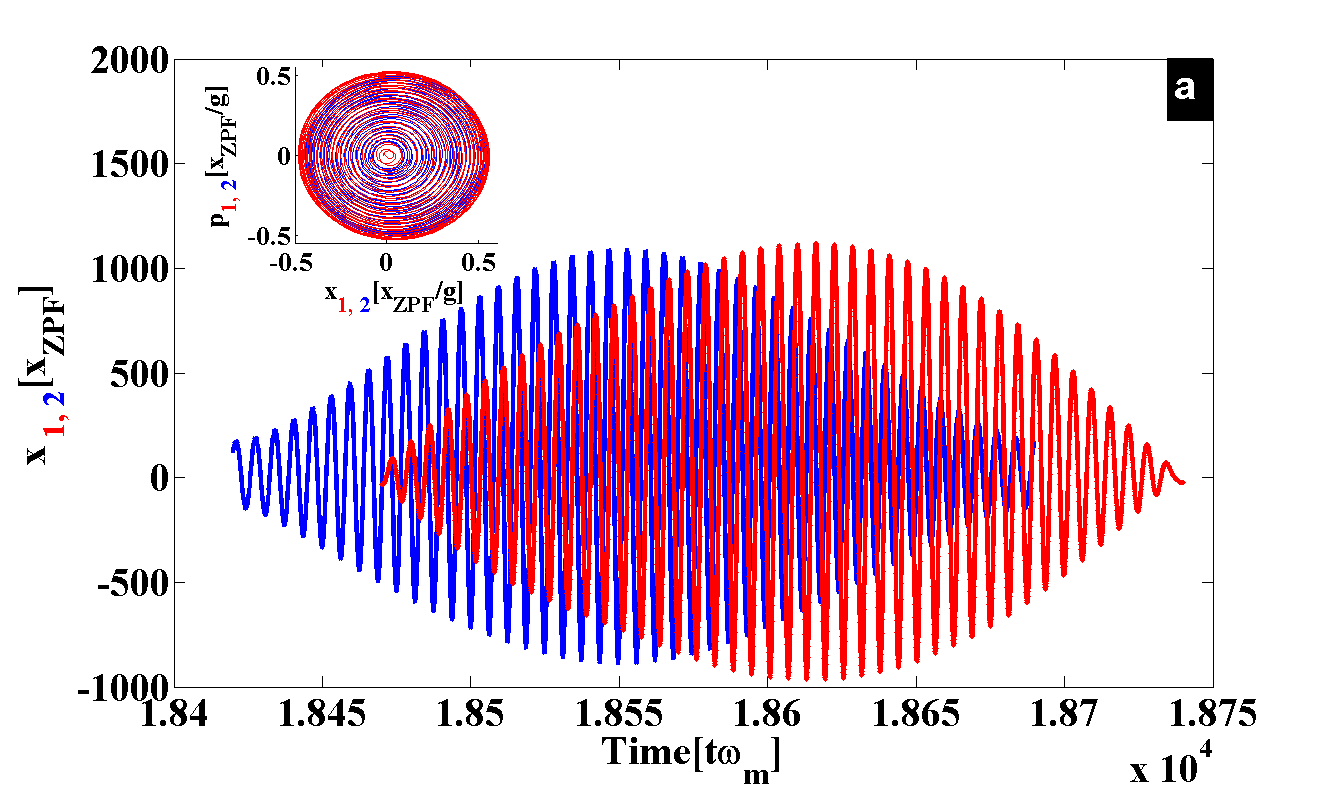}}
\resizebox{0.38\textwidth}{!}{
\includegraphics{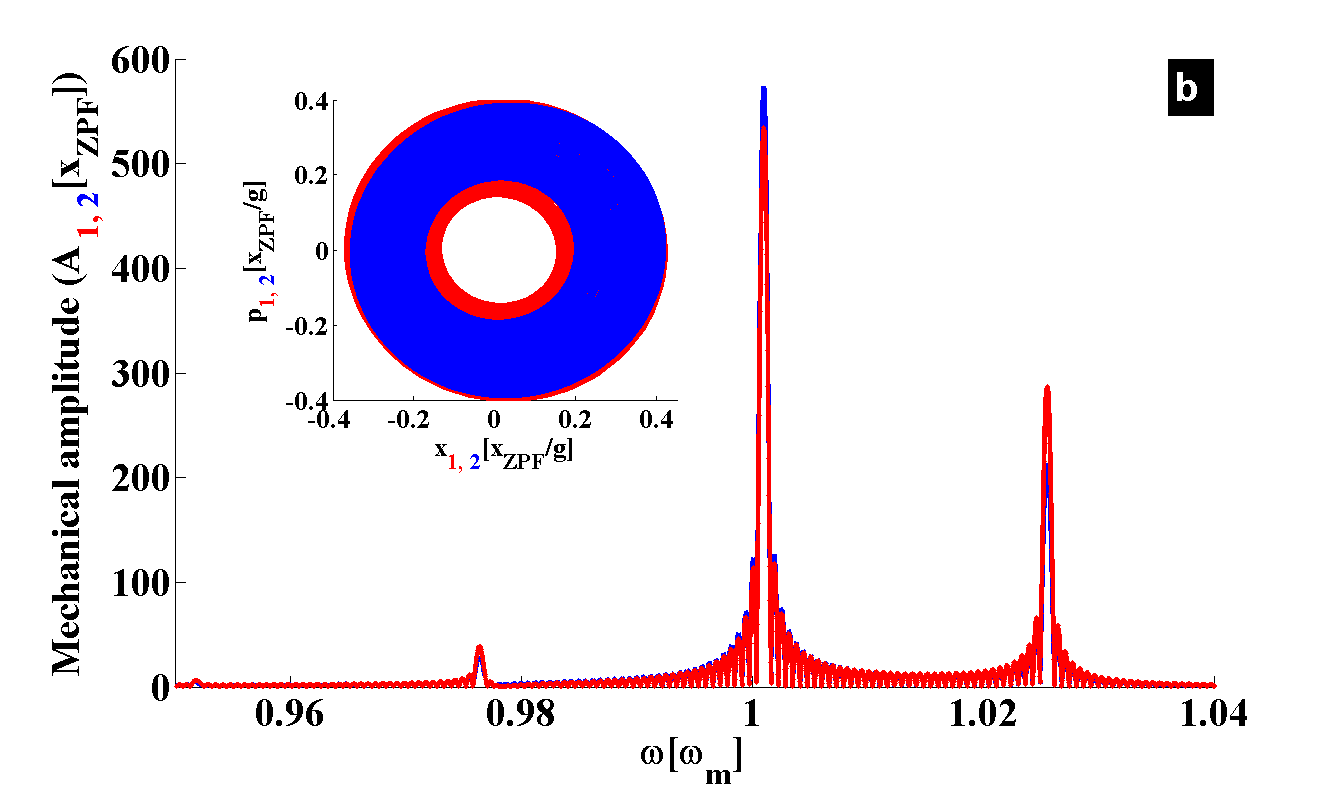}}
\resizebox{0.38\textwidth}{!}{
\includegraphics{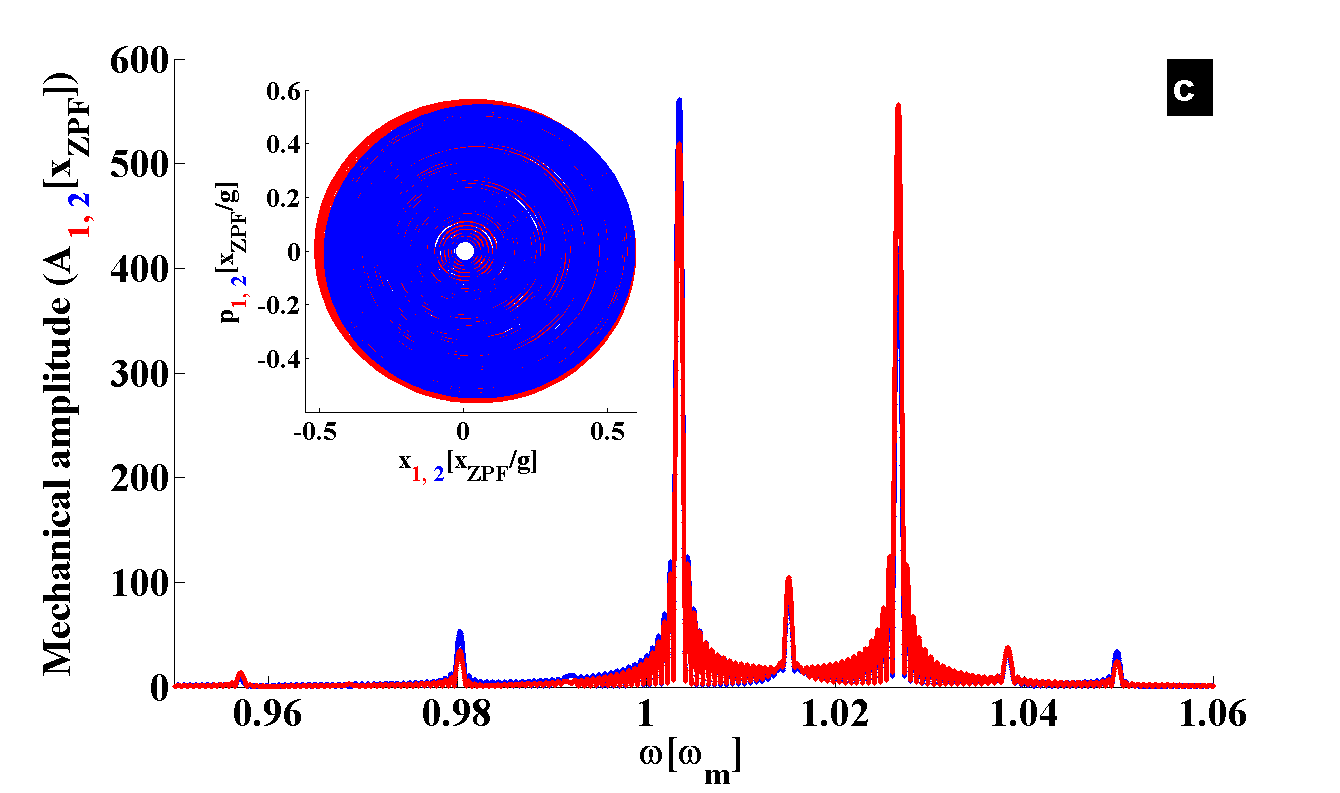}}
\resizebox{0.38\textwidth}{!}{
\includegraphics{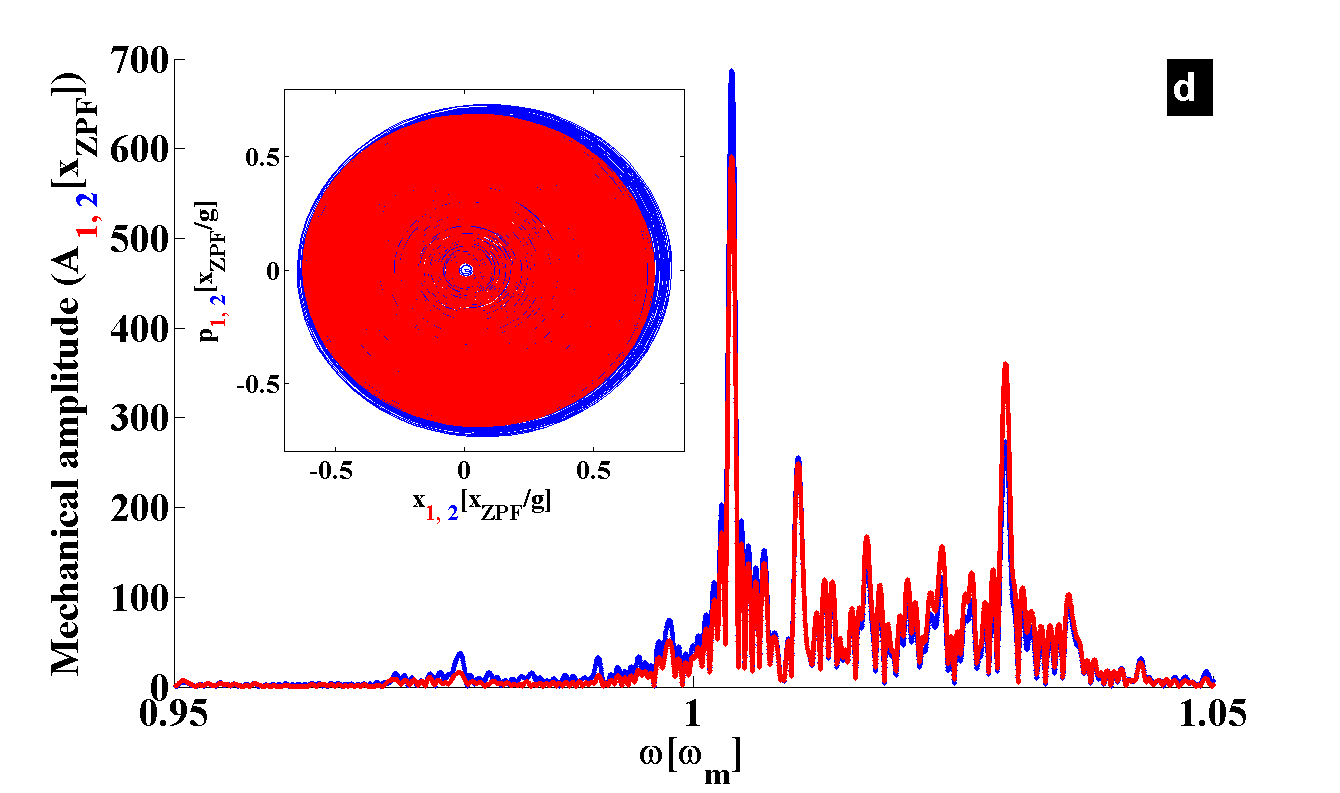}}
\end{center}
\caption{Fourier spectra and phase space trajectories. (a) One Rabi cycle and the corresponding phase space, featuring 
a set of limit cycles. (b) Regular Rabi oscillations at $\alpha^{in}= 4.5\times10^2\sqrt{\omega_m}$. The corresponding 
time propagation is Fig.1d in the main text. (c) Quasi-periodic state corresponding to 
Fig. \ref{fig:FigS1}c, some satellite peaks can be observed. (d) Chaotic state corresponding to 
Fig. \ref{fig:FigS1}d, several peaks have emerged. Insets of these figures are their phase space trajectories,
which all feature a set of limit cycles. Consequently, Fourier spectra are useful to distinguish these states, 
instead of the phase space representations. Blue (red) color is related to 
the blue (red) mechanical supermode.}
\label{fig:FigS2}
\end{figure*}

Numerical steady state solutions of Eq. (\ref{eq1}) are shown by Fig.1b in the 
main text. Three dynamical states are depicted, the fixed point regime, the limit cycles regime, and  
the regime where Rabi oscillations emerge. The aim  here is to characterize dynamically, 
the steady states solutions in these regimes. For this purpose, we have fixed $J=2.2\times10^{-2}\omega_m$,
where all these regimes are met by varying the driving strength $\alpha^{in}$ (see the horizontal dashed line in Fig.1b
in the main text). Hence, time propagation of some steady state solutions are given in Fig. \ref{fig:FigS1}. 
Fixed point regime is shown in Fig. \ref{fig:FigS1}a, where Rabi oscillations 
are decaying with a same rate as explained in the main text. The frequency 
of these Rabi oscillations is $\sigma_{0}$, and the mechanical resonators are 
in strong coupling regime. Fig. \ref{fig:FigS1}b represents limit cycle oscillations 
at $\alpha^{in} = 4.3\times10^{2} \sqrt{\omega_m}$, and the system is in a weak coupling regime. 
Figs. \ref{fig:FigS1}(c)$-$(d), show Rabi oscillations in the nonlinear regime 
(see green area in Fig.1b in the main text). At $\alpha^{in} = 5\times10^{2} \sqrt{\omega_m}$,  
 Fig. \ref{fig:FigS1}c shows quasi-periodic behaviour, and at less three  frequencies can be observed. 
In Fig. \ref{fig:FigS1}d however, several frequencies are involved, resulting in chaotic dynamics.
To further characterize these dynamical states, we have used bifurcation diagram in frequency 
space (see Fig.3b in the main text). The reason lies on the difficulty of catching
Rabi oscillation's dynamics in phase space. Indeed, Fig. \ref{fig:FigS2}a shows one period of 
Rabi oscillations with its corresponding phase space representation in the inset.
Accordingly, this phase space trajectory features a set of limit cycles, each of those corresponds
to each amplitude involved in the Rabi cycle \cite{[40]}. Figs. \ref{fig:FigS2}(b)$-$(d) are 
the Fourier spectra corresponding respectively to regular ($\alpha^{in} = 4.5\times10^{2} \sqrt{\omega_m}$), 
quasi-periodic ($\alpha^{in} = 5\times10^{2} \sqrt{\omega_m}$) and 
chaotic Rabi oscillations ($\alpha^{in} = 5.5\times10^{2} \sqrt{\omega_m}$). The insets of these 
figures are the phase space representations, and they all feature a set of limit cycles 
as in Fig. \ref{fig:FigS2}a. It follows that, phase space representation is not a useful tool 
to distinguish between different dynamical states here. However, Fourier spectra in 
Figs. \ref{fig:FigS2}(b)$-$(d) clearly discriminate the dynamical states involved. Hence, 
varying the driving $\alpha^{in}$ at a fixed $J=2.2\times10^{-2}\omega_m$, we were able to 
construct Fig.3b of the main text from Fast Fourier Transform (FFT).

\section{Analytics}  \label{App.B}

In the limit cycles regime, the amplitudes of the mechanical oscillations change
only slowly over time (see Fig. \ref{fig:FigS1}b). Thus, we  solve the equation 
for $\alpha_{j}$ assuming a fixed amplitude for the mechanical oscillations, and then use the
result to solve the equation for $\beta_{j}$ \cite{[41]}, \cite{[35]}. Under this
assumption, the mechanical oscillation can be described by the ansatz,

\begin{equation}
\beta_{j}(t)=\bar{\beta}_{j}+A_{j}\exp(-i\omega_{lock}t)\text{,}
\label{S1}
\end{equation}%
where $\bar{\beta}_{j}$ is a constant shift in the origin of the resonator
and the amplitude $A_{j}$ is taken to be a slowly varying function of time. 
In such a weak coupling regime, we have denoted the locked frequency by $\omega_{lock}$.
We substitute this ansatz into the equation for $\alpha_{j}$, and use the
assumption of a slowly evolving amplitude to solve it, first neglecting the
time dependence of $A_{j}$ \cite{[35]}, \cite{[41]}. We then obtain the intracavity field
in the form,

\begin{equation}
\alpha_{j}(t)=e^{-i\theta_{j}\left( t\right) }\sum_{n}\alpha
_{n}^{j}e^{in \omega_{lock}t}\text{.}  \label{S2}
\end{equation}
The phase is $\theta_{j}\left( t\right) =-\epsilon_{j}\sin \omega_{lock}t$
and the amplitudes of the different harmonics of the optical field are,
\begin{equation}
\alpha_{n}^{j}=-i\sqrt{\kappa_{j}}\alpha ^{in}\frac{J_{n}\left(
-\epsilon_{j}\right) }{h_{n}^{j}}\text{,}  \label{S3}
\end{equation}
where $\epsilon_{j}=\frac{2g\rm{Re}(A_{j})}{\omega_{lock}}$ , $\tilde{%
\Delta}_{j}=\Delta_{j}+2g\rm{Re}(\bar{\beta}_{j})$, $h_{n}^{j}=i\left(
n\omega_{lock}-\tilde{\Delta}_{j}\right) +\frac{\kappa }{2}$ and $J_{n}$ is
the Bessel function of the first kind of order $n$.

As we are interested in the regime of limit cycles of the resonators, a
rotating wave approximation can be made in which we drop all the terms (in
the mechanical dynamics) except the constant one and the term oscillating at
$\omega _{lock}$. Hence, we substitute Eq. (\ref{S2}) in the equation for $%
\beta _{j}$ (see Eq. (\ref{S1})) which, by equating constant terms, leads
to the zero-frequency components,

\begin{equation}
\left\{
\begin{array}{c}
\bar{\beta}_{1}=\frac{1}{\omega_{01}-i\frac{\gamma_{m}}{2}}\left(g\kappa
\sum_{n}\frac{\left(\alpha^{in}J_{n}\left(-\epsilon_{1}\right)
\right)^{2}}{\left\vert h_{n}^{1}\right\vert ^{2}}+J\bar{\beta}_{2}\right),
\\
\bar{\beta}_{2}=\frac{1}{\omega_{02}-i\frac{\gamma_{m}}{2}}\left(g\kappa
\sum_{n}\frac{\left(\alpha^{in}J_{n}\left(-\epsilon_{2}\right)
\right)^{2}}{\left\vert h_{n}^{2}\right\vert ^{2}}+J\bar{\beta}_{1}\right),
\end{array}
\right.  \label{S4}
\end{equation}
that induce a shifts of the cavity frequencies,
\begin{equation}
\delta_{j}=2g\rm{Re}(\bar{\beta}_{j})\text{.}  \label{S5}
\end{equation}

The equations of motion for the oscillating part of $\beta _{j}$ are deduced
from $\beta_{r}^{j}(t)=\beta_{j}(t)-\bar{\beta}_{j}\equiv A_{j}\exp
(-i\omega_{lock}t)$ and read,
\begin{equation}
\left\{
\begin{array}{c}
\dot{\beta}_{r}^{1}(t)=-i\left(\omega_{01}+\delta \omega_{1}\right) \beta
_{r}^{1}-\frac{\gamma_{m}+\gamma_{opt}^{1}}{2}\beta_{r}^{1}+iJ\beta
_{r}^{2} \\
\dot{\beta}_{r}^{2}(t)=-i\left(\omega_{02}+\delta \omega_{2}\right) \beta
_{r}^{2}-\frac{\gamma_{m}+\gamma_{opt}^{2}}{2}\beta_{r}^{2}+iJ\beta
_{r}^{1}
\end{array}
\right.  \label{S6}
\end{equation}

Here the optical spring effect $\delta \omega_{j}$ and the optical damping $%
\gamma_{opt}^{j}$ coming both from the average dynamics of the cavity are
given by,

\begin{equation}
\delta \omega_{j}=-\frac{2\kappa (g\alpha^{in})^{2}}{\omega
_{lock}\epsilon_{j}}\rm{Re}\left(\sum_{n}\frac{J_{n+1}\left(
-\epsilon_{j}\right) J_{n}\left(-\epsilon_{j}\right)}{%
h_{n+1}^{j\ast }h_{n}^{j}}\right) \text{,}  \label{S7}
\end{equation}
and

\begin{equation}
\gamma_{opt}^{j}=\frac{2(g\kappa \alpha^{in})^{2}}{\epsilon_{j}}%
\sum_{n}\frac{J_{n+1}\left(-\epsilon_{j}\right) J_{n}\left(
-\epsilon_{j}\right)}{\left\vert h_{n+1}^{j\ast }h_{n}^{j}\right\vert
^{2}}\text{.}  \label{S8}
\end{equation}

For $\epsilon_{j}\ll 1$, the linear approximation is still valid and both the optical
spring effect and the optical damping can be rewritten accordingly. Indeed, $%
\epsilon_{j}\ll 1$ induces $J_{n}\left(-\epsilon_{j}\right) \approx
\frac{1}{n!}\left(\frac{-\epsilon_{j}}{2}\right)^{n}$ for $n\geq 0$
and $J_{-n}\left(-\epsilon_{j}\right) =J_{n}\left(\epsilon
_{j}\right)$. Using these considerations in Eq. (\ref{S7}) and Eq. (\ref{S8}) yield,
\begin{widetext}

\begin{equation}
\delta \omega_{j}\left(0\right) \approx -\frac{2\left(g_{j}\alpha
_{j}^{in}\right) ^{2}\kappa_{j}\tilde{\Delta}_{j}\left[ \frac{3\kappa
_{j}^{2}}{4}+\left(\omega_{lock}-\tilde{\Delta}_{j}\right) \left(\omega_{lock}+%
\tilde{\Delta}_{j}\right) \right] }{\left(\frac{\kappa_{j}^{2}}{4}\left[
\frac{\kappa_{j}^{2}}{4}+\left(\omega_{lock}-\tilde{\Delta}_{j}\right)
\left(\omega_{lock}+\tilde{\Delta}_{j}\right) \right] -\kappa_{j}\tilde{%
\Delta}_{j}^{2}\right)^{2}+\tilde{\Delta}_{j}^{2}\left[ \frac{3\kappa
_{j}^{2}}{4}+\left(\omega_{lock}-\tilde{\Delta}_{j}\right) \left(\omega_{lock}+%
\tilde{\Delta}_{j}\right) \right]^{2}},  \label{S07}
\end{equation}

\begin{equation}
\gamma_{opt}^{j}\left(0\right) \approx -\frac{\tilde{\Delta}_{j}\omega
_{lock}\left(2g_{j}\kappa_{j}\alpha_{j}^{in}\right)^{2}}{\left(\tilde{%
\Delta}_{j}^{2}+\frac{\kappa_{j}^{2}}{4}\right) \left[ \left( \omega_{lock}+%
\tilde{\Delta}_{j}\right) ^{2}+\frac{\kappa _{j}^{2}}{4}\right] \left[
\left(\omega_{lock}-\tilde{\Delta}_{j}\right) ^{2}+\frac{\kappa_{j}^{2}}{4}%
\right]}\text{.}  \label{S08}
\end{equation}

\end{widetext}

These expressions are well in agreement with what is obtained in the linear
regime \cite{[41]}, where both $\delta \omega_{j}$ and $\gamma_{opt}^{j}$ are 
not amplitude dependent.

\section{Effective Hamiltonian}  \label{App.C}

From Eq.(\ref{S6}), it is possible to define effective Hamiltonian in order
to figure out supermodes involved in the system. Such supermodes will be
deduced from the eigenmodes of the effective model, describing the mechanical resonators. 
Indeed, the real parts of the eigenmodes give the eigenfrequencies of the coupled system while their
imaginary parts stand for the dissipations rate of the system. In the limit cycles
regime, the constant shift $\bar{\beta}_{j}$ is weak compared to the
amplitude of the mechanical resonator ($\bar{\beta}_{j}\ll A_{j}$).
This means that $\beta_{j}(t)\cong \beta_{r}^{j}(t)$, and Eq.(\ref{S6})
can be assumed as a set of equations describing the effective system that
reads,

\begin{equation}
\left\{
\begin{array}{c}
\dot{\beta}_{1}=-\left(i\omega_{eff}^{1}+\frac{\gamma_{eff}^{1}}{2}%
\right) \beta_{1}+iJ\beta_{2}, \\
\dot{\beta}_{2}=-\left(i\omega_{eff}^{2}+\frac{\gamma_{eff}^{2}}{2}%
\right) \beta_{2}+iJ\beta_{1},
\end{array}
\right.  \label{S10}
\end{equation}
where $\omega_{eff}^{j}=\omega_{0j}+\delta \omega_{j}$ and $\gamma
_{eff}^{j}=\gamma_{m}\pm \gamma_{opt}^{j}$ define the effective
frequencies and the effective damping, respectively.

Furthermore, Eq.(\ref{S10}) can be rewritten in the compact form,

\begin{equation}
\partial t\Psi =-iH_{eff}\Psi
\end{equation}
with the effective Hamiltonian,
\begin{equation}
H_{eff}=
\begin{bmatrix}
\omega_{eff}^{1}-i\frac{\gamma_{eff}^{1}}{2} & -J \\
-J & \omega_{eff}^{2}-i\frac{\gamma_{eff}^{2}}{2}
\end{bmatrix}
\label{S11}
\end{equation}
and the state vector $\Psi =\left(\beta_{1},\beta_{2}\right)^{T}$.

The eigenvalues of the Hamiltonian given in Eq.(\ref{S11}) are obtained by
solving the equation,

\begin{equation}
\det \left( H_{eff}-\lambda I\right) =0,
\end{equation}
and that yields to the following eigenvalues $\lambda_{-}$ and $\lambda_{+}$,
\begin{equation}
\lambda_{\pm}\simeq  \frac{\omega_{eff}^{1}+\omega_{eff}^{2}}{2}-\frac{i}{4}\left(\gamma _{eff}^{1}+\gamma
_{eff}^{2}\right) \pm \frac{\sigma}{2}.  \label{S12}
\end{equation}
with $\sigma =\sqrt{4J^{2}-\frac{\Delta \gamma_{eff}^{2}}{4}}$ and $\Delta
\gamma_{eff}=\gamma_{eff}^{1}-\gamma_{eff}^{2}$. The frequencies and the
dissipations of the supermodes are given by the real and imaginary parts of $%
\lambda_{\pm}$, respectively

\begin{equation*}
\omega_{\pm}=\rm{Re} \left(\lambda_{\pm}\right) \text{ and }\gamma
_{\pm }=\rm{Im} \left(\lambda_{\pm}\right) \text{.}
\end{equation*}

From Eq.(\ref{S12}), we deduce whether the system is in strong coupling regime
or not. Indeed, for $J>\frac{\Delta \gamma_{eff}}{4}$,  $\sigma$ is real and this induces two distinct
frequencies,

\begin{equation}
\omega_{\pm}=\frac{\omega_{eff}^{1}+\omega_{eff}^{2}}{2} \pm \frac{\sigma}{2}
\end{equation}
that are spectrally separated by
\begin{equation}
\sigma =\omega_{+}-\omega_{-}=\sqrt{4J^{2}-\frac{\Delta \gamma_{eff}^{2}}{%
4}}\text{.}  \label{S13}
\end{equation}
This splitting modes is the sign of strong coupling between the resonators
and  $\sigma$ is the frequency of Rabi oscillations that emerge. The mechanical
resonators have the same damping $\gamma_{\pm }=\frac{-\left(
\gamma_{eff}^{1}+\gamma_{eff}^{2}\right)}{4}$. However, for $%
J<\frac{\Delta \gamma_{eff}}{4}$, $\sigma$ is imaginary and the resonators 
oscillate at the same frequency,
\begin{equation}
\omega_{\pm}=\frac{\omega_{eff}^{1}+\omega_{eff}^{2}}{2},
\end{equation}
with two distinct dissipations,
\begin{equation}
\gamma_{\pm}=\frac{-\left(\gamma_{eff}^{1}+\gamma
_{eff}^{2}\right)}{4}\pm \frac{\sigma }{2}\text{.}
\end{equation}
This corresponds to a regime where the mechanical resonators are weakly coupled. 
The phase transition  between these two regimes happens at the exceptional 
point (\textbf{EP}), where $J=\frac{\Delta\gamma_{eff}}{4}$ that is equivalent to 
 $\sigma = 0$. To demonstrate the feature of multiple \textbf{EPs}, we need to 
 show that $\sigma = 0$ can leads to multiple solutions. For this purpose, 
 let us remind that  $\Delta \gamma_{eff}\equiv\Delta \gamma_{opt}^{j}$. 
 Then, using Eq.(\ref{S8}) in Eq.(\ref{S13}), leads straightforwardly to understand 
 that $\sigma$ is amplitude-dependent through the Bessel functions. Due to these Bessel functions, 
 $\sigma = 0$ oscillates.  These oscillations of $\sigma$, depending on the 
 system's parameters, induce multiple solutions of $\sigma = 0$, resulting in multiple \textbf{EPs} feature.

\end{document}